\documentclass[journal=jpcafh,manuscript=article]{achemso}

\usepackage{graphicx,color}
\usepackage{amsmath,amsfonts,amssymb}
\usepackage{mathtools}
\usepackage{dcolumn}
\usepackage{lmodern}
\usepackage{comment}
\usepackage[normalem]{ulem}
\usepackage{cancel}
\usepackage{xspace}
\usepackage{mhchem}

\newcommand{\fig}{Fig.}
\newcommand{\Fig}{Fig.}
\newcommand{\figref}[1]{\fig~\ref{#1}}
\newcommand{\Figref}[1]{\Fig~\ref{#1}}

\renewcommand{\eqref}[1]{Eq.~(\ref{#1})}
\newcommand{\Eqref}[1]{Eq.~(\ref{#1})}

\newcommand{\new}[1]{#1}
\newcommand{\neww}[1]{#1}

\newcommand{\rem}[1]{}%\textcolor{red}{\sout{#1}}}
\newcommand{\stkout}[1]{\ifmmode\text{\sout{\ensuremath{#1}}}\else\sout{#1}\fi}

\newcommand{\braketop}[3]{\ensuremath{\left\langle#1\middle|#2\middle|#3\right\rangle}}
\newcommand{\ii}{\text{i}}   
\newcommand{\ee}{\text{e}} 
\newcommand{\unit}{\mathbf{I}}

\title{Calculation of Reaction Rate Constants in the Canonical and Microcanonical Ensemble}
\author{Andreas L\"{o}hle}
\author{Johannes K\"{a}stner}
\email{kaestner@theochem.uni-stuttgart.de}
\affiliation{Institute for Theoretical Chemistry, University of Stuttgart, 
		Pfaffenwaldring 55, 70569 Stuttgart,Germany}
\date{}

\begin{document}
\begin{abstract}
  Canonical instanton theory is a widespread approach to describe the dynamics
  of chemical reactions in low temperature environments when tunneling
  effects become dominant. It is a semiclassical theory which requires
  locating classical periodic orbits on the upside-down potential energy
  surface, so-called instantons, and the computation of second order quantum
  corrections. The calculation of these corrections usually involves a matrix
  diagonalization. In this paper we present \new{an alternative approach,} which requires to
  solve only linear systems of equations involving sparse
  matrices. Furthermore the proposed method provides a reliable and
  numerically stable way to obtain stability parameters in multidimensional
  systems, which are of particular interest in the context of microcanonical
  instanton theory.
	
\end{abstract}

%	\affiliation{Institute for Theoretical Chemistry, University of Stuttgart,
%		Pfaffenwaldring 55, 70569 Stuttgart,Germany, kaestner@theochem.uni-stuttgart.de}
	
%\date{\today}

%	\pacs{}
	%\keywords{}
	
	\maketitle

%%%%%%%%%%%%%%%%%%%%%%%%%%%%%%%%%%%%%%%%%%%%%%%%%%%%%%%%%%%%%%%%%%%%%%%%%%
\section{Introduction}
%%%%%%%%%%%%%%%%%%%%%%%%%%%%%%%%%%%%%%%%%%%%%%%%%%%%%%%%%%%%%%%%%%%%%%%%%%

The phenomenon of tunneling is a significant feature of almost any quantum
mechanical system.  In the context of chemistry the effects of quantum
tunneling are particularly important when it comes to rate constants for
chemical reactions that take place in cold environments, for instance in gas
clouds in interstellar space, or involve the transfer of light atoms such as
hydrogen.\cite{miy04,koh05,all09,koh03,nag06,bor16,mei16} While an exact
quantum mechanical treatment would be most desirable in order to asses
tunneling contributions it is usually not achievable due to the high
computational effort which is required for realistic and hence more complex
higher dimensional systems.\cite{fer06,pu06,nym14,kae14} Over the years
various methods have been suggested in order to handle these problems. A
promising and nowadays widespread approach is what is usually referred to as
canonical instanton theory. It is in essence a semiclassical theory which has
appeared in somewhat different formulations since the
1960's. \cite{lan67,lan69,mil75, col77,cal77} A common approach to introduce
instanton theory is the imaginary-$F$ premise which relates the free energy $F$
of the system to the thermal reaction rate constant $k$ as\cite{cal77, aff81,
  ben94}
\begin{align}
  k(\beta) = - \frac{2}{\hbar} \text{Im } F \approx \frac{2}{\hbar \beta}
  \frac{\text{Im } Q}{\text{Re } Q}  \label{haha0}
\end{align}
whereby $Q$ is the system's canonical partition function and $\beta$ is the
inverse temperature, $\beta=1/(k_\text{B}T)$. The task of calculating
$k(\beta)$ is therefore reduced to finding the real and imaginary part of the
partition function whereby the real part represents the partition function of
the reactant state $Q_\text{RS}$ and the imaginary part the partition function
of the transition state $Q_\text{TS}$. By using a well known analogy between
the Feynman path integral representation of the Schr\"odinger propagator in
quantum mechanics and the calculation of the partition function in
statistical physics one obtains a path integral representation in imaginary
time\cite{Feynman_book}  for the partition function.
\begin{align}
  Q &= \int d\mathbf{x} \braketop{\mathbf{x}}{\ee^{-\beta \hat{H}}}{\mathbf{x}} \\
  &= \oint \mathcal{D}\mathbf{x}(\tau) \ee^{-\mathcal{S}_\text{E}[x(\tau)]} \label{g3}
\end{align}
with the Euclidean action functional given by 
\begin{equation}
\mathcal{S}_\text{E}[x(\tau)] =
\int \frac{M \dot{x}(\tau)^2}{2} + V(x(\tau)) d\tau
\end{equation}
and a Hamiltonian of the form $\hat{H} = P^2/(2M) + V(x)$. The path integral
in \eqref{g3} is then approximated by the action of the classical solutions,
which give the dominant contributions to the path integral, and the effect of
deviations $\delta \mathbf{x}$ from the classical path. The canonical
partition function, using this semiclassical approximation, can therefore be
written as
\begin{align}
  Q^\text{SC} = \sum_i F_i \  \ee^{-\mathcal{S}^i_\text{cl}} 
\end{align}
where $\mathcal{S}^i_\text{cl}$ is the Euclidean action of the $i^{\text{th}}$
classical solution and $F_i$ is the fluctuation factor that contains quantum
corrections up to second order.  The classical trajectories are solutions to
the classical equation of motion in imaginary time $m \ddot{\mathbf{x}} =
\nabla V$. The sign change of the potential is a consequence of the 
Wick rotation from real to imaginary time. Usually two classical solutions can
be found. One is a particle resting still at $\mathbf{x}(\tau) =
\mathbf{x}_\text{RS}$ corresponding to $Q_\text{RS}$, the fluctuation factor
$F_\text{RS}$ of which is entirely real. A second solution, the so-called
instanton, is an unstable periodic orbit with the period $\beta\hbar$ and an
entirely imaginary fluctuation factor $F_\text{Inst}$. The rate constant is
then given by
\begin{align}
  k(\beta) = \frac{2}{\hbar \beta} \frac{F_\text{Inst}}{F_\text{RS}} \ee ^{-(\mathcal{S}_\text{Inst} - \mathcal{S}_\text{RS})/\hbar}. \label{haha}
\end{align}
To locate an instanton, the path is discretized into $P$ points (images or
replicas of the system) and stationary points of the discretized Euclidean
action functional are searched for, rather than integrating the classical
equation of motion with the right initial conditions. For a system with $D$
spatial degrees of freedom the actional function becomes a function of $DP$
variables. The instanton search can nowadays efficiently be done by using a
truncated Newton search.\cite{arn07,and09,rom11,rom11b} For the fluctuation
factor this discretization scheme yields a multidimensional Gaussian integral
of the form
\begin{align}
  F = \int_{-\infty}^\infty \exp{\left(-\frac 1 2 \sum\limits_{i,j=1}^{DP}A_{ij} x_i x_j \right)} \, d^{DP}x  = \sqrt{\frac{(2\pi)^{DP}}{\text{det}\mathbf{A}}} \label{soso}
\end{align}
where $\mathbf{A}$ is $P D \times P D$ matrix which contains the Hessians of
the potential along all $P$ discrete points of the instanton trajectory.\cite{rom11b} The
final result on the right hand side of \eqref{soso} is only applicable if
$\mathbf{A}$ is positive definite. In the case of the instanton, which is a
saddle point of the discretized action functional, this is not the case as it
has one negative eigenvalue and at least one zero eigenvalue. However, there
exist ways to deal with these eigenvalues in a physically sensible manner and
render $F$ finite.\cite{FADDEEV67,kle09} The conventional way of obtaining $F$
is therefore to obtain the full eigenvalue spectrum of
$\mathbf{A}$.\cite{col88,han90, ben94,mes95,arn07,ric09, and09,kry11,alt11,
  rom11,rom11b,kae14,ric18} We will refer to this method from now on as the
determinant method. Since diagonalizing a matrix is one of the most demanding
tasks in terms of the computational effort required, a faster approach is
desirable. In this paper we are going to present a method that allows a much
faster calculation of $F$.

Rather than using the straightforward way of calculating the discretized path
integral for the canonical partition function by solving a product of
Gaussian integrals, we start in the microcanonical ensemble and, via a steepest
descent approximation, obtain an expression for the fluctuation
factor. \new{The approach bears some similarities to previous work,\cite{ric16,ric16a,ric18a} however
it avoids the construction of a coordinate system which separates
reactive and orthogonal modes along the instanton path.} It only requires solving two linear systems of equations.

The paper is organized as follows. First we briefly review the basics of
microcanonical instanton theory as well as the definitions of the monodromy
matrix and the Van-Fleck propagator. Then we present the necessary algorithms
to compute the quantities previously derived in order to compute $F$. Finally
we test the new methods on two different systems, the analytic M\"uller--Brown
potential and a chemical reaction with $V(x)$ and its derivatives calculated
on-the-fly by density functional theory. In all cases the results are compared
to the conventional determinant method.  Finally we discuss the advantages and
disadvantages of the new method.

%%%%%%%%%%%%%%%%%%%%%%%%%%%%%%%%%%%%%%%%%%%%%%%%%%%%%%%%%%%%%%%%%%%%%%%%%% 
\section{Theory}
%%%%%%%%%%%%%%%%%%%%%%%%%%%%%%%%%%%%%%%%%%%%%%%%%%%%%%%%%%%%%%%%%%%%%%%%%% 
\label{sec::theory}
\subsection{Microcanonical Instanton Theory} \label{MTT}

The key quantity of microcanonical instanton theory is the so called
cumulative reaction probability $P(E)$,\cite{mil75} which is the result of
averaging all state-to-state cross sections for a system at a fixed energy
$E$. The microcanonical rate constant $k(E)$ can then be calculated as follows
\cite{mil91}
\begin{align}  
  k(E) = \frac{1}{2 \pi \hbar} \frac{P(E)}{\Gamma_r(E)}
\end{align} 
with $\Gamma_r(E)$ being the density of states of the reactant. An efficient
way to obtain $P(E)$ directly is the use of the quantum flux-flux
autocorrelation formalism \cite{mil98a,mil98add} which gives an exact expression for
$P(E)$
\begin{align}
  P(E) = 2 \pi^2 \hbar^2 \text{tr}\left(\delta(E-\hat{H}) \hat{F} \delta(E-\hat{H})
    \hat{F} \right) \label{final}
\end{align}	 
where $\delta(E-\hat{H})$ is the density operator in the microcanonical
ensemble and $\hat{F}$ is the quantum mechanical analogue of the classical
flux function which counts the number of elementary reactions from reactant to
product and is given by
\begin{align}
  \hat{F} = \frac{\text{i}}{\hbar} \left[\hat{H}, \hat{\theta} (s) \right].
\end{align}
Here, $\hat{\theta}$ is the Heaviside step function and $s$ denotes a
parameter that is negative on the reactant side of a dividing surface and
positive on the product side. In one dimension, $P(E)$ can be obtained by a
variety of methods, including a direct numerical solution of Schr\"odinger's
equation. Instanton theory provides a semiclassical approximation 
for $P(E)$ for a system with $D_\nu$ vibrational degrees of freedom as:\cite{mil75}
\begin{eqnarray}
%\begin{multline}
  P_\text{SC}(E)&=&\sum_{k=1}^\infty(-1)^{k-1}\prod_{i=1}^{D_\nu-1}\frac{1}{2\sinh(ku_i(E)/2)}\times\nonumber\\
  &&\times \exp(-k\mathcal{S}_0(E))  \\
 &=& \sum_{k=1}^\infty(-1)^{k-1} F_{k,\perp}(E) \exp(-k\mathcal{S}_0(E)).
  \label{eq1}
\end{eqnarray}
%\end{multline}
Here and in the following, atomic units
($\hbar=e=m_\text{e}=4\pi\epsilon_0=1$) and mass weighted coordinates are
used, $\mathcal{S}_0 = \int \dot{\mathbf{x}}^2/2 d\tau$ is the shortened action, and
$u_i(E)$ are the stability parameters of the instanton path. The corresponding
thermal rate constant can be obtained by thermally averaging $k(E)$ which
results in a Laplace transform of $P(E)$. 
	 %EQUATION 3
%\begin{align}
%  k(\beta) = \frac{1}{2 \pi Q_{\text{RS}}} \int_{-\infty}^{\infty} P(E)
%  \exp(-\beta E) dE. \label{PEW}
%\end{align}
If we use \eqref{eq1} as approximation for $P(E)$ we assume the rotational
motion to be separable from the internal motion. \Eqref{eq1} is essentially
$P(E,J)$ for $J=0$. In the $J$-shifting approximation\cite{tak52,bow91} the
rotational dependence is taken out of the integral such that the thermal rate
constant is given by
\begin{align}
  k(\beta) = \frac{Q_\text{t-r}}{2 \pi Q_{\text{RS}}}
  \int_{-\infty}^{\infty} P_\text{SC}(E) \exp(-\beta E) dE \label{PEW}
\end{align}
where $Q_\text{t-r}$ is the ratio of translational and rotational partition
functions of transition state and reactant. Rotational partition functions are
typically estimated from rigid rotors.

In principle one has to sum over all classical solutions with the same energy
$E$, which also includes solutions that pass the instanton's orbit multiple
times. Since the contributions of those trajectories are weighted with
$\exp{(-k \mathcal{S}_0)}$, with $k$ being the multiplicity of the orbit, their
contributions decay exponentially and can be neglected. However, for an
instanton with an energy $E$ close to the transition state's energy
$E_\text{TS}$ this becomes an increasingly bad approximation since close to
$E_\text{TS}$ the shortened action $\mathcal{S}_0$ gets smaller and ultimately vanishes
at $E = E_\text{TS}$. This neglect of contributions from repetitions of the
instanton orbit leads to the familiar overestimation of $k(\beta)$ close to
the crossover temperature $T_c$. Techniques to correct for that were
proposed.\cite{kry13,mcconnel17} In order to perform the Laplace transform in
\eqref{PEW} we approximate the integral for $k=1$ via a steepest descent
approach such that
\begin{align}
  \int_{-\infty}^{\infty} F_\perp(E)  \ee^{-\left( \beta E + \mathcal{S}_0\right)} dE \approx \sqrt{\frac{{2 \pi}}{\frac{d^2\mathcal{S}_0}{dE^2}}} F_\perp(E_0) \ee^{-\left( \beta E_0 + \mathcal{S}_0(E_0)\right)} 
\end{align}
where $E_0$ satisfies the condition
$\frac{d}{dE} \left( \mathcal{S}_0 + \beta E \right) = 0$. From that we obtain 
$\frac{d \mathcal{S}_0}{d E} = -\beta$ which results in\cite{mil75} 
\begin{align}
  k(\beta) = \frac{Q_\text{t-r}}{\sqrt{2 \pi}
  Q_\text{RS} } \sqrt{-\frac{dE}{d\beta}}
  \prod_{i=1}^{D_\nu-1}\frac{1}{2\sinh(u_i(E_0)/2)}
  \ee^{-\mathcal{S}_\text{Inst}} \label{eq10}
\end{align}
with $\mathcal{S}_\text{Inst}=\mathcal{S}_\text{E}(E_0)$.
So once an instanton is located $\frac{dE}{d\beta}$  and the stability parameters $u_i$ 
have to be determined in order to
calculate the rate constant.
%vibrational partition function of the transition
%state. We assume a decoupling of rotation and translation for reactant
%and transition state such that
%	 \begin{align}
%	 Q_\text{TS}  &= Q_\text{Inst} \cdot  Q_\text{Rot} \cdot Q_\text{Trans} \\
%	 Q_\text{RS}  &= Q^{\text{RS}}_\text{Vib} \cdot  Q^{\text{RS}}_\text{Rot} \cdot Q^{\text{RS}}_\text{Trans} 
%	 \end{align}
 %    For the rotational and translational partition functions the classical expressions are used in both cases.

\subsection{The Van-Fleck Propagator and the Monodromy Matrix}
\label{mono_theo}	

To determine the stability parameters we have to
find a way to compute the monodromy matrix \cite{gut71} first.  For
that purpose we will need to evaluate the Van-Fleck propagator which
is a semiclassical expression for the Schr\"odinger propagator.  In
imaginary time it is given by \cite{gut67, ric15a}
\begin{align}
  K_\text{sc}(\mathbf{x}'' ,\mathbf{x}' ,\ii t) = \left(\frac{1}{2 \pi }\right)^{\frac{D}{2}} \sqrt{ \left| \text{Det} 
  \left(-\frac{\partial^2 \mathcal{S}_\text{E}}{\partial x'_i \partial x''_j } \right) \right|}  e^{-\mathcal{S}_\text{E} + \ii \Phi} \label{VF}
\end{align}
whereby $\Phi$ is 
\begin{align}
  \Phi = -\frac{\pi}{2} \nu
\end{align}
and $\nu$ is called the Maslov--Morse index which counts the number of
zeros of the determinant. \Eqref{VF} gives the semiclassical probability amplitude of
a particle with unit mass moving from $\mathbf{x}'$ to $\mathbf{x}''$
in imaginary time $\tau = \ii t$. To calculate the
partition function we have to take the trace of \eqref{VF}
\begin{align}
Q^\text{SC} =  \int    \ K_\text{sc}(\mathbf{x} ,\mathbf{x} ,\beta) d\mathbf{x} 
\end{align}    
 In order to evaluate the trace we perform a steepest descent integration. Using that for the classical solution $\frac{\partial \mathcal{S}_\text{E}}{\partial \mathbf{x}} = 0$ and 
 \begin{align}
 \frac{\partial^2 \mathcal{S}_\text{E}}{\partial \mathbf{x}^2} = \left(  \frac{\partial^2 \mathcal{S}_\text{E}}{\partial \mathbf{x'} \mathbf{x}'} + 2 \frac{\partial^2 \mathcal{S}_\text{E}}{\partial \mathbf{x}'\mathbf{x}''} +  \frac{\partial^2 \mathcal{S}_\text{E}}{\partial \mathbf{x}'' \mathbf{x}''} \right)\Big|_{\mathbf{x'} = \mathbf{x}'' = \mathbf{x}}
\end{align}
we finally get
\begin{align}
Q^{\text{SC}} = \sum_i Q^{\text{SC},i}
\end{align}
with 
\begin{equation}
  \begin{split}
    Q^{\text{SC},i}  =&\sqrt{ \frac{\left|-\frac{\partial^2 \mathcal{S}^i_\text{E}}{\partial \mathbf{x}' \partial \mathbf{x}'' } \right|_{\substack{x' = x''=x}}}{ \left| \frac{\partial^2 \mathcal{S}^i_\text{E}}{\partial \mathbf{x}' \partial \mathbf{x}' } + 2 \frac{\partial^2 \mathcal{S}^i_\text{E}}{\partial \mathbf{x}' \partial \mathbf{x}'' } + \frac{\partial^2 \mathcal{S}^i_\text{E}}{\partial \mathbf{x}'' \partial \mathbf{x}'' } \right|_{\substack{x' = x''=x}}}} \times \\ &\exp\left( -\mathcal{S}^i_{\text{cl}} -\ii \frac{\pi}{2} \nu  _i \right) \\
    &= F_i  \exp\left( -\mathcal{S}^i_{\text{cl}} \right) \exp\left(  -\ii \frac{\pi}{2} \nu_i  \right)  \label{eq2} 
  \end{split}	
\end{equation}
Since the classical solution that corresponds to the reactant state is simply
a particle resting still, it has no turning points, hence $\nu = 0$ and the
Euclidean action simply becomes $\mathcal{S}_{\text{RS}} = \beta
V(\mathbf{x}_\text{RS})$. The instanton is a closed orbit and
therefore $\nu$ depends on how often the particle reaches the turning
points. In this case $\nu$ can have values of $2k$ where $k$ gives the number
of repetitions of the instanton orbit. However, we have to an add an
additional phase factor of $-\ii \pi/2$ in order to account for the fact that
the instanton travels only in the classically forbidden region and therefore
only contributes to the imaginary part of the partition function. If we consider one orbit only for the instanton we
get $\nu = 2$ and the fluctuation factor turns imaginary as $\exp({-\ii ( \nu
  \frac{\pi}{2}} + \frac{\pi}{2})) = \ii$. This gives for the partition
functions of the reactant and the transition states
\begin{align}
  Q_\text{RS} &= F_\text{RS} \ee^{-\mathcal{S}_\text{RS}} \\
  Q_\text{Inst} &= \ii F_\text{Inst} \ee^{-\mathcal{S}_\text{Inst}}.
\end{align}
The term $F$ can in principle be divergent. This is due to symmetries
present in the system which lead to an over-counting of real physical
states and therefore to a diverging partition function. Since the
instanton is a closed orbit the action is invariant with respect to
the choice of the starting and end point as
$\mathbf{x}' = \mathbf{x}'' = \mathbf{x}$. Furthermore rotation and
translational invariance are also symmetries that lead to diverging
terms, yet these can easily be handled as those symmetries are also
present in the reactant partition function and therefore cancel one
another if one is only interested in their ratios. In order to handle
the divergent terms we express $F$ in a different representation. We
construct a matrix $\mathbf{M}$ with the following entries
\begin{align}
\mathbf{M} = \begin{pmatrix}
-\mathbf{b}^{-1} \mathbf{a} &  -\mathbf{b}^{-1} \\
\mathbf{b} - \mathbf{c} \mathbf{b}^{-1} \mathbf{a}  &  -\mathbf{c} \mathbf{b}^{-1} 
\end{pmatrix} \label{MM45}
\end{align}
whereby $\mathbf{a}, \mathbf{b}$ and $\mathbf{c}$ are defined as
\begin{align}
 \mathbf{a}  = \frac{\partial \mathcal{S}_\text{E}}{\partial \mathbf{x}' \partial \mathbf{x}'} \Big|_{\mathbf{x'} = \mathbf{x}'' = \mathbf{x}}  \\
\mathbf{b} = \frac{\partial \mathcal{S}_\text{E}}{\partial \mathbf{x}' \partial \mathbf{x}''} \Big|_{\mathbf{x'} = \mathbf{x}'' = \mathbf{x}}  \\
\mathbf{c} = \frac{\partial \mathcal{S}_\text{E}}{\partial \mathbf{x}''\partial \mathbf{x}''} \Big|_{\mathbf{x'} = \mathbf{x}'' = \mathbf{x}} 
\end{align}
such that the fluctuation factor in \eqref{eq2} can be written as 
\begin{align}
F = \sqrt{\frac{ \left| - \mathbf{b}\right|}{\left| \mathbf{a} + 2 \mathbf{b} + \mathbf{c}\right|}} = \sqrt{\frac{(-1)^D}{\left| \mathbf{M} - \mathbf{1} \right|}}. \label{mm11}
\end{align}
%Now if we evaluate the right hand site of \eqref{mm11} and choose to 
If $\mathbf{M}$ is represented in its eigenbasis one can immediately
see that the right-hand side of
\eqref{mm11} diverges as $\mathbf{M}$ has eigenvalues of $\lambda =
1$. These correspond to the symmetries of the system. Furthermore
$\mathbf{M}$ is a symplectic matrix meaning that eigenvalues always appear in
pairs. If $\lambda_i$ is an eigenvalue so is $1/\lambda_i$. The matrix
$\mathbf{M}$ is called the monodromy matrix.\cite{gut71} It contains
information about how significantly a solution to the classical equations of
motion deviates from its reference position and momentum under an
infinitesimal perturbation of $\delta \mathbf{x}_0$ and $\delta \mathbf{p}_0$
after one period $T_0$ has passed.
  \begin{align}
\begin{pmatrix}
 \delta \mathbf{x} \\
 \delta \mathbf{p}
 \end{pmatrix} = \mathbf{M}(T_0)
  \begin{pmatrix}
 \delta \mathbf{x}_0 \\
 \delta \mathbf{p}_0
 \end{pmatrix}  
  \end{align}
  where $\mathbf{M}$ is the solution to the  linearized equations of motion 
\begin{align}
\frac{d}{dt}\mathbf{M}(t) = \begin{pmatrix}
 0 & \mathbf{1} \\
 - \mathbf{V}''\left((\mathbf{x}\left(t\right)\right) & 0
 \end{pmatrix} \mathbf{M}(t)  \label{mono}
\end{align}
evaluated at $t = T_0$. Thus, the eigenvalues of $\mathbf{M}$ inform
about the stability of the trajectory. Real values of $\lambda$
are connected to unstable modes. With the definition 
$\lambda_i = e ^{ u_i}$ we get the fluctuation factor
\begin{align}
F &= \sqrt{\prod_{i=1}^{D} \frac{-1}{(e^{u_i} - 1) \cdot (e^{-u_i}- 1 )}}\\
              &= \sqrt{\prod_{i=1}^{D} \frac{-1}{2 - 2 \cosh(u_i)}} \\
              &=  \prod_{i=1}^{D} \frac{1}{2 \sinh{(u_i/2)}}.\label{sinh}
\end{align}  
The fluctuation factor $F$ including all $D$ degrees of freedom is obviously
divergent as it has at least one, in the case of rotational and translational
symmetries an additional six, zero-valued stability parameters $u_i = 0$. As
mentioned before we can simply ignore the translational and rotational ones as
the fluctuation factor only covers vibrations. Using \eqref{haha}, gives for
the rate constant
\begin{align}
  k(\beta) = \frac{2}{\beta} Q_\text{t-r} \frac{
    2\prod_{j=1}^{D_\nu} \sinh(u_j^\text{RS}/2) }{\prod_{i=1}^{D_\nu-1}
    \sinh(u_i^\text{Inst}/2) }
  F^\text{Inst}_{\parallel} \ee
  ^{-\mathcal{S}_\text{Inst} + \mathcal{S}_\text{RS}}. \label{finaleq}
\end{align}
The parallel fluctuation factor $F^\text{Inst}_{\parallel}$, which only
appears in the instanton partition function, can not be described by
\eqref{sinh} due to $u_i$ being zero.  However, it can be addressed by
applying the Faddeev--Popov trick to avoid the over-counting of ghost states
in order to render the partition function finite.\cite{FADDEEV67} In this case,
however, we can simply compare \eqref{haha} with \eqref{eq10}, using
$F^\text{Inst}=F^\text{Inst}_{\perp}F^\text{Inst}_{\parallel}$, and obtain
\begin{align}
  \frac{2}{\beta} \frac{Q_\text{t-r} Q_\text{Inst}}{Q_\text{RS} } &= \frac{Q_\text{t-r}}{\sqrt{2 \pi} Q_\text{RS} }	\sqrt{-\frac{dE}{d\beta}} F^\text{Inst}_{\perp} \ee^{-\mathcal{S}_\text{Inst}} \notag\\
  \frac{2}{\beta} F^\text{Inst} \ee^{-\mathcal{S}_\text{Inst}} &= \frac{1}{\sqrt{2\pi}} \sqrt{-\frac{dE}{d\beta}} F^\text{Inst}_{\perp} \ee^{-\mathcal{S}_\text{Inst}} \notag \\
  F^\text{Inst}_{\parallel} &= {\sqrt{\frac{\beta^2}{8 \pi}}
    \sqrt{-\frac{dE}{d\beta}}}. \label{haha35}
\end{align}
Using the result of \eqref{haha35} we obtain the final rate constant expression
\begin{align}
  k(\beta) = \frac{Q_\text{t-r}}{\sqrt{ 2\pi}}\sqrt{-\frac{dE}{d\beta}}
  \frac{ 2\prod_{j=1}^{D_\nu} \sinh(u_j^\text{RS}/2) }{\prod_{i=1}^{D_\nu-1}
    \sinh(u_i^\text{Inst}/2) } \ee ^{-\mathcal{S}_\text{Inst} +
    \mathcal{S}_\text{RS}}.\label{final_final}
\end{align} 
Overall, in order to evaluate the thermal rate constant we find an
instanton at given $\beta$ and calculate its stability parameters
$u_i$, as well as $\frac{dE}{d\beta}$. In a previous paper
\cite{mcconnel17} we have presented several methods using different
approximative schemes to calculate $u_i$ as an alternative to the
traditional way of integrating the linearized equations of motions,
\eqref{mono}.  Here, we present a different approach making use of
\eqref{MM45}, which allows us to calculate the monodromy matrix by
using the second variations of $\mathcal{S}_\text{E}$. It is exact in
the limit of $P \to \infty$ but appears to be numerically more stable
than integrating \eqref{mono}. In case of the reactant state,
$u_i^\text{RS}\equiv \omega_i^\text{RS}\beta$ with
$\omega_i^\text{RS}$ being the reactant's vibrational frequencies, see
Appendix. However, in order to benefit from error cancellation at
finite $P$, we use the same numerical algorithm for the reactant state
as we do for the instanton. In the following, we present a numerical
algorithm to evaluate $\frac{dE}{d\beta}$ and $u_i$.

\section{Numerical implementation}\label{num_imp}

In the following sections we present two algorithms in order to
calculate the necessary quantities mentioned in the previous
sections. In both cases we rely on a discretization of the classical
equation of motion in imaginary time, i.e. in the upside-down
potential $-V$, which yields
\begin{align}
\ddot{\mathbf{x}}(\tau) &= \nabla V \\
\left(-2 \mathbf{x}_i + \mathbf{x}_{i+1} + \mathbf{x}_{i-1} \right) \frac{1}{\Delta \tau^2} &= \nabla V(\mathbf{x_i}) \\
\left(2 \mathbf{x}_i - \mathbf{x}_{i+1} - \mathbf{x}_{i-1} \right)  + \frac{\beta^2}{P^2} \nabla V(\mathbf{x_i})  &= 0. \label{3}
\end{align} 
The vector $\mathbf{x}_i$ is a $D$ dimensional array that contains the
positions for every degree of freedom at the $i^\text{th}$ image. $P$
is the number of discrete points and $\beta$ is in this context the
period in imaginary time such that $\Delta \tau = \frac{\beta}{P}$.

Furthermore it is worth mentioning that the kind of instanton solutions that
we use in the following have the property that each image appears exactly
twice as the spatial coordinates of the trajectory from the reactant to the
product side are the same as on the way back from product to reactant. This
leads to $\mathbf{x}_i = \mathbf{x}_{P-i+1}$. However, the following
derivation does not use this fact and the algorithms are valid for
any closed trajectory which may or may not posses turning points.

\subsection{Calculation of $\frac{dE}{d\beta}$} \label{dedb} 

In order to calculate the change of the energy of the instanton
solution with respect to $\beta$ we start by looking at the energy
conservation of the instanton solution.\neww{\cite{alt11,ric15b}} Since we are in imaginary time
the momentum in the classically forbidden region is purely imaginary
and therefore the energy $E$ is given by 
\begin{align}
E &= \frac{\ii ^2 \mathbf{p}^2}{2} + V(\mathbf{x}) \\
  &= - \frac{\mathbf{p}^2}{2} + V(\mathbf{x}). \label{eq7}
\end{align}
It is conserved along the orbit. The discretized form of \eqref{eq7} reads
\begin{align}
  E = \lim_{P \to \infty }\left( - \frac{ \left( \mathbf{x}_i -
        \mathbf{x}_{i-1} \right)^2}{2} \left(\frac{P}{\beta}\right)^2 +
  \frac 1 2 (  V(\mathbf{x}_i)+ V(\mathbf{x}_{i-1}))\right) \label{eq8}
\end{align}
Since a change in $\beta$ means that a new instanton solution for a new
temperature has to be found we can interpret the points $\mathbf{x}_i$
along the trajectory to be a function of $\beta$, thus
$\mathbf{x}_i = \mathbf{x}_i(\beta)$. If we now differentiate
\eqref{eq8} with respect to $\beta$ we get
\begin{align}
\frac{dE}{d\beta} =& \lim_{P \to \infty }  \Bigg( \frac{ P^2}{\beta^3} \left(\mathbf{x}_i - \mathbf{x}_{i-1}\right)^2 \notag \\ &- \frac{P^2}{\beta^2}\left(\mathbf{x}_i - \mathbf{x}_{i-1}\right) \left( \frac{d\mathbf{x}_i}{d\beta} - \frac{d\mathbf{x}_{i-1}}{d\beta} \right)  \notag \\
 &+ \frac 1 2 \left(
 \frac{\partial V}{\partial \mathbf{x}_i} \frac{d\mathbf{x}_i}{d\beta} +
 \frac{\partial V}{\partial \mathbf{x}_{i-1}} \frac{d\mathbf{x}_{i-1}}{d\beta} 
\right) \Bigg) \label{eq9}
\end{align}
In order to evaluate \eqref{eq9} one has to calculate
$\frac{d\mathbf{x}_i}{d\beta}$. The first step is to differentiate
\eqref{3} with respect to $\beta$ and replacing
$\frac{d\mathbf{x}_i}{d\beta}$ by $\mathbf{q}_i$. This yields\neww{\cite{ric15b}}
\begin{align}
\left( 2 + \frac{\beta^2}{P^2} \nabla^2 V(\mathbf{x}_i) \right) \mathbf{q}_i -\mathbf{q}_{i+1} -\mathbf{q}_{i-1} = - \frac{2 \beta}{P^2} \nabla  V(\mathbf{x}_i) \label{5}
\end{align}
\Eqref{5} is  a linear system of equations of the form 
\begin{align}
\mathbf{A}\mathbf{q} = \mathbf{b} \label{5.5}
\end{align} 
or, using an index notation for images and degrees of freedom,
\begin{align}
\sum_{i'= 1}^P \sum_{j' = 1}^{D} A_{i, j, i', j'} \ q_{i', j'} = b_{i, j}.
\end{align}
Here,
\begin{align}
\mathbf{A} = 
 \begin{pmatrix}
\mathbf{K}_1  & -\unit  & 0 &   \cdots & 0 & -\unit  \\
-\unit & \ddots  & -\unit &   \ddots & \ddots & 0  \\
0 & -\unit  &  \mathbf{K}_i&   \ddots & \ddots & \vdots  \\
\vdots & \ddots  & \ddots & \mathbf{K}_{i+1} & -\unit & 0  \\
0 & \ddots & \ddots &   -\unit & \ddots  & -\unit   \\
-\unit & 0  & \cdots &   0 & -\unit & \mathbf{K}_P
 \end{pmatrix} 
\end{align}
where $\mathbf{K}_i = 2 \unit + \frac{\beta^2}{P^2}
\mathbf{V}''(\mathbf{x}_i)$, $\unit$ being the $D\times D$ unit matrix and
%\begin{align}
%b_{i, j} = - \frac{2 \beta}{P^2} \nabla_j  V(\mathbf{x}_i)
%\end{align}
\begin{align}
\mathbf{b} = - \frac{2 \beta}{P^2} 
\begin{pmatrix}
\nabla_1  V(\mathbf{x}_1) \\
\vdots\\
\nabla_D  V(\mathbf{x}_1) \\
\nabla_1  V(\mathbf{x}_2) \\
\vdots \\
\nabla_j  V(\mathbf{x}_i) \\
\vdots \\
\nabla_D  V(\mathbf{x}_P)
\end{pmatrix}.
\end{align}
Solving \eqref{5.5} requires the knowledge of all Hessians as well as
gradients along the instanton path. The matrix $\mathbf{A}$ is a symmetric
sparse $\left(PD \times PD \right)$ matrix.

Having obtained $\frac{d\mathbf{x}_i}{d\beta}=\mathbf{q}_i$ we can now
determine $\frac{dE}{d\beta}$ by choosing any arbitrary point $\mathbf{x}_i$
and use it in \eqref{eq9}. In practice the value of $\frac{dE}{d\beta}$ will
slightly vary along the path due to numerical errors. Therefore, we choose one
of the turning points as the distance between the current and the next point
is the shortest in that case.

\subsection{Calculation of $\frac{\partial \mathcal{S}_\text{E}}{\partial
    \mathbf{x}' \partial \mathbf{x}''}$} \label{vara}

Since an analytical treatment of the second derivatives of
$\mathcal{S}_\text{E}$ is only possible in the case of a separable potential
in which the orthogonal components of the potential have a linear or quadratic
form, we have to find a way to determine these necessary expressions
numerically, \neww{which is achieved by following previous work. \cite{alt11,ric15b}} We start
with the discretized Euclidean action for an instanton that starts at $\tau =
0$ at the point $ \mathbf{x}' = \mathbf{x}_0$ and ends at $\tau = \beta$ at
the point $\mathbf{x}'' = \mathbf{x}_P$. Furthermore a constant time interval
$\Delta \tau = \beta / P$ is used because we consider an arbitrary open path
with $P+1$ images to derive the first and second variations and only then
close the path by setting $\mathbf{x}' = \mathbf{x}'' = \mathbf{x}$ resulting
in $P$ independent images. The Euclidean action is
\begin{align}
  \mathcal{S}_\text{E} = \sum_{i = 1}^{P} \left[ \frac{1}{2} \frac{(\mathbf{x}_i - \mathbf{x}_{i-1})^2}{\Delta\tau} + \frac{\Delta \tau}{2} \left( V(\mathbf{x}_i) + V(\mathbf{x}_{i-1})\right) \right] \label{v1}
\end{align}
The first variation of \eqref{v1} results in
\begin{align}
\frac{\partial \mathcal{S}_\text{E}}{\partial \mathbf{x}'} &= - \frac{(\mathbf{x}_1 - \mathbf{x}_0)}{\Delta \tau} + \frac{\Delta\tau}{2} V'(\mathbf{x}_0) \\
\frac{\partial \mathcal{S}_\text{E}}{\partial \mathbf{x}''} &=  \frac{(\mathbf{x}_P - \mathbf{x}_{P-1})}{\Delta \tau} + \frac{\Delta\tau}{2} V'(\mathbf{x}_P) 
\end{align}
To calculate the second variation of \eqref{v1} we have to keep in mind that
while $\mathbf{x}'$ and $\mathbf{x}''$ stay fixed, the points in between
change. In this case we regard $\mathbf{x}_i$ for $i \in [1, ..., P-1]$ as a
function of $\mathbf{x}_0$ and $\mathbf{x}_P$. This results in the second
variation
\begin{align}
\frac{\partial^2 \mathcal{S}_\text{E}}{\partial \mathbf{x}' \partial \mathbf{x}''} &= 
- \frac{1}{\Delta \tau} \frac{\partial \mathbf{x}_1}{\partial \mathbf{x}_P}  
%=- \frac{1}{\Delta \tau} \frac{\partial \mathbf{x}_{P-1}}{\partial \mathbf{x}_0}  
\label{hj5} \\
\frac{\partial^2 \mathcal{S}_\text{E}}{\partial \mathbf{x}' \partial \mathbf{x}'} &=  \frac{1}{\Delta \tau} \left( \unit- \frac{\partial \mathbf{x}_1}{\partial \mathbf{x}_0} \right) + \frac{\Delta \tau}{2} \mathbf{V}''(\mathbf{x}_0)  \label{hj6}\\
\frac{\partial^2 \mathcal{S}_\text{E}}{\partial \mathbf{x}'' \partial \mathbf{x}''} &=  \frac{1}{\Delta \tau} \left( \unit - \frac{\partial \mathbf{x}_{P-1}}{\partial \mathbf{x}_P} \right) + \frac{\Delta \tau}{2} \mathbf{V}''(\mathbf{x}_P) \label{hj7}
\end{align}
To be able to use these expressions we have to find a way to calculate the
terms $\frac{\partial \mathbf{x}_{P-1}}{\partial \mathbf{x}_P}$ and
$\frac{\partial \mathbf{x}_2}{\partial \mathbf{x}_1}$. This can be done in the
same way as before based on \eqref{3}. We differentiate \eqref{3} with respect
to $\mathbf{x}_\alpha$ where $\alpha$ can be either $0$ or $P$ and get\neww{\cite{ric15b}}
\begin{align}
 2 \frac{ \partial \mathbf{x}_{i}}{\partial \mathbf{x}_\alpha} - \frac{ \partial \mathbf{x}_{i+1}}{\partial \mathbf{x}_\alpha} - \frac{ \partial \mathbf{x}_{i-1}}{\partial \mathbf{x}_\alpha} + \Delta \tau^2 \mathbf{V}''(\mathbf{x_i}) \frac{ \partial \mathbf{x}_{i}}{\partial \mathbf{x}_\alpha} &= 0 \label{gh45}\\
\left(  2 \unit + \Delta \tau^2 \mathbf{V}''(\mathbf{x_i}) \right) \mathbf{J}_i - \mathbf{J}_{i+1} -  \mathbf{J}_{i-1} &= 0 \\
\mathbf{K}_i \mathbf{J}_i - \mathbf{J}_{i+1} -  \mathbf{J}_{i-1} &= 0
\end{align}
where $i \in \left[ 1, ..., P-1 \right]$ and $\mathbf{J}_i \equiv \frac{ \partial \mathbf{x}_{i}}{\partial \mathbf{x}_\alpha}$ is a $D \times D$ matrix with the following boundary conditions for $\alpha = 1$
\begin{align}
\mathbf{J}_{i = 0}^{\alpha = 0} = \unit \\
\mathbf{J}_{i = P}^{\alpha = 0} = 0 
\end{align}
and for $\alpha = P$
\begin{align}
\mathbf{J}_{i = 0}^{\alpha = P}  = 0 \\
\mathbf{J}_{i = P}^{\alpha = P} = \unit
\end{align}
In principle what we have obtained here is another representation of
\eqref{mono}. If we take \eqref{gh45}, which is a second order differential
equation of the form $\ddot{\mathbf{J}}(\tau) = -
\mathbf{V}''(\mathbf{x}(\tau)) \mathbf{J}(\tau)$, and transform it to first
order we get the familiar matrix differential equation of \eqref{mono}.  As
solving \eqref{mono} with a Runge--Kutta approach turned out rather unstable
in practice\cite{mcconnel17} for low values of $P$ we will instead focus on
solving \eqref{gh45}. First we transform \eqref{gh45} to a linear system of
equations of the form
\begin{align}
\mathbf{C} \mathbf{q} = \mathbf{d} \label{foh4}
\end{align} 
in which $\mathbf{C}$ is matrix of dimension $D^2 (P-1) \times D^2 (P-1)$,
which is given as
\begin{align}
  \mathbf{C} = 
  \begin{pmatrix}
    \mathbf{G}_1  & -\unit  & 0 &   \cdots & \cdots & 0 \\
    -\unit & \ddots  & -\unit &   \ddots & \cdots & 0  \\
    0 & -\unit &  \mathbf{G}_i&   \ddots & \ddots & \vdots  \\
    \vdots & \ddots  & \ddots & \mathbf{G}_{i+1} & \ddots  & \vdots  \\
    \vdots & \ddots & \ddots &   \ddots & \ddots  & -\unit  \\
    0 & 0  & \cdots &   \cdots & -\unit& \mathbf{G}_{P-1}
  \end{pmatrix} 
\end{align}
with $\unit$ being a $D^2 \times D^2$-dimensional unit matrix and
$\mathbf{G}_i$ is a $D^2 \times D^2$ matrix of the form
\begin{align}
  \mathbf{G}_i =  \begin{pmatrix}
    \mathbf{K}_i  &  & 0 \\
    & \ddots  &    \\
    0 &   &  \mathbf{K}_i& 
  \end{pmatrix} .
\end{align}
The right hand side of \eqref{foh4} is given by the column vectors of
$\mathbf{J}_0$ or $\mathbf{J}_P$ depending on whether we want to calculate
$\frac{\partial }{\partial \mathbf{x}_P}$ or $\frac{\partial}{\partial \mathbf{x}_0}$. For example in the case of a system
with $D = 2$, $\mathbf{d}$ is given by
\begin{align}
d = \begin{pmatrix}
1 \\
0 \\
0 \\
1 \\
0 \\
\vdots \\
0
\end{pmatrix}
\text{ for } \frac{\partial}{\partial \mathbf{x}_0},
% \ \ \ \ \ \ \ \ \  \ \ 
d = \begin{pmatrix}
0 \\
\vdots \\
0 \\
1 \\
0 \\
0 \\
1
\end{pmatrix}
\text{ for } \frac{\partial}{\partial \mathbf{x}_P}.
\end{align}
Correspondingly the solutions we are looking for are given by
\begin{align}
  \frac{\partial \mathbf{x}_1}{\partial \mathbf{x}_0} = \begin{pmatrix}
    q_1 & q_3\\
    q_2 & q_4 
  \end{pmatrix} \ \ \ \ \ \ \ \frac{\partial \mathbf{x}_{P-1}}{\partial \mathbf{x}_P} = \begin{pmatrix}
    q_{DP-4} & q_{DP-2}\\
    q_{DP-3} & q_{DP-1} 
  \end{pmatrix}.
\end{align}
This scheme works in general for any path whether it is open or
closed. In the closed case one just sets $\mathbf{x}_0 = \mathbf{x}_P$ in
Eqs. (\ref{hj5}), (\ref{hj6}), and (\ref{hj7}). Instead of calculating the
determinant of a ($DP \times DP $) matrix one now has to solve a linear system
of equations of the form $\mathbf{C}\mathbf{q} = \mathbf{d}$, where
$\mathbf{C}$ is a banded matrix of the size $D^2 (P-1)$ . 

In summary we have to evaluate \eqref{final_final} in order to compute the
thermal rate constant $k(\beta)$.  The vibrational partition function of the reactant
and the orthogonal contributions to the transition state's
partition function $Q_\text{Inst}$ are calculated via the
stability parameters, which are obtained from the eigenvalues of the monodromy
matrix $\mathbf{M}$ obtained by \eqref{MM45}. The entries of $\mathbf{M}$ are
calculated by Eqs.~(\ref{hj5}) to (\ref{hj7}), which require solving the
linear system of equations in \eqref{foh4}. For that, the Hessians
($\mathbf{V}''(\mathbf{x}_i)$) of all images along the path have to be
known. The derivative $dE/d\beta$ is calculated by solving \eqref{5.5} and
using the solution in \eqref{eq9}.

% \jk{timings oder Diskussion D vs. P -- temporary data, to be removed later.}
% Diagonalization scales with $N^3$

% Measured scaling for 4 to 64 images on JK's laptop (and desktop for
% comparions), system blas and MKL give the same scaling. MKL is overall faster,
% though.

% dE/dbeta scales $P^3$, so does the full diagonalization (factor 30-40
% more expensive, though). $u_i$-computation scales linearly with $P$.

% $dE/d\beta$ linear system, matrix size $DP$, scaling measured
% $\mathcal{O}(P^3)$, scaling must be $\mathcal{O}(P^3D^3)$ \jk{is that true?
%   can the D-dependence be different?}

% full diagonalization, matrix size $DP$, scaling measured
% $\mathcal{O}(P^3)$, scaling must be $\mathcal{O}(P^3D^3)$.

% $u_i$-calculation, linear system (band matrix), matrix size $D^2(P-1)$,
% measured scaling $\mathcal{O}(P)$ $\mathcal{O}(D^6)$ 

% \jk{end of temporary stuff}

The computational requirements of this algorithm are different from the
determinant method. In the latter, one sparse but full matrix of size $DP$
needs to be diagonalized, which requires $\mathcal{O}(P^3D^3)$ operations. In
our algorithm, the calculation of $dE/d\beta$, i.e. solving \eqref{5.5},
requires to solve a sparse linear system of $DP$ equations, which again
requires $\mathcal{O}(P^3D^3)$ operations. The pre-factor is 1--2 orders of
magnitude lower, however. Thus, the calculations are faster by that factor. To
obtain the $u_i$-values, a system of equations of the size $D^2(P-1)$ has to
be solved, \eqref{foh4}. The matrix is a banded matrix with a width of only
$\mathcal{O}(D^2)$, thus requiring $\mathcal{O}(PD^6)$
operations. Consequently, our algorithm is faster than the determinant method
in practice, especially for small systems with many images, which is typically
the case for fitted potential energy surfaces.

%\jk{we need some name or label for that method: variational (action) method,
 % derivative method, ...}

\section{Applications} \label{MBS}

 We applied the new method to two examples: the analytic, two-dimensional
M\"uller--Brown potential and a rearrangement of methylhydroxycarbene to
acetaldehyde with energies, gradients and Hessians calculated on-the-fly by
DFT.

\subsection{M\"uller--Brown Potential}

\begin{figure}[htbp!]
\includegraphics[width=8cm]{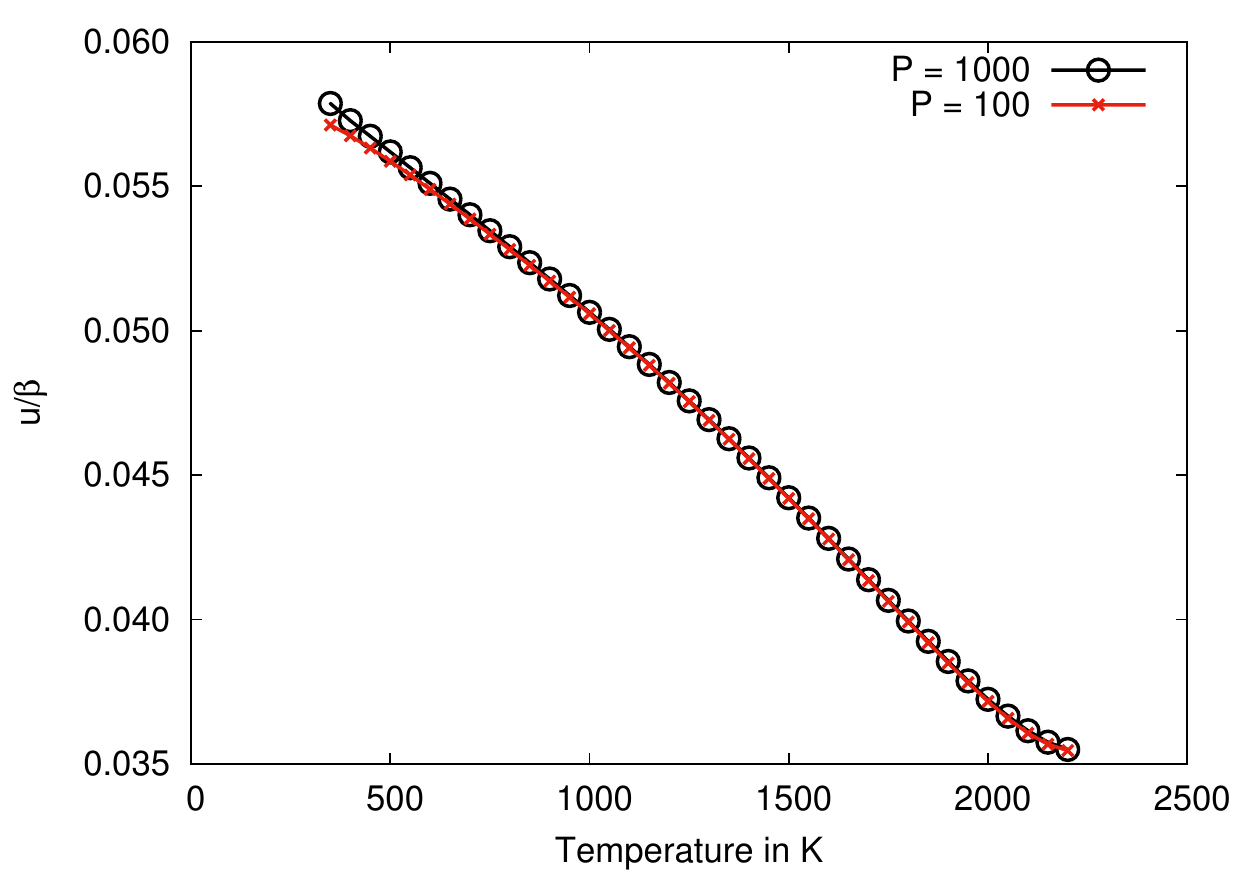}
\caption{Stability parameter of the instanton's orthogonal modes in the
  M\"uller--Brown potential calculated at different temperatures for
  different numbers of images $P$.}
\label{pic11}
\end{figure}

\begin{figure}
\includegraphics[width=8cm]{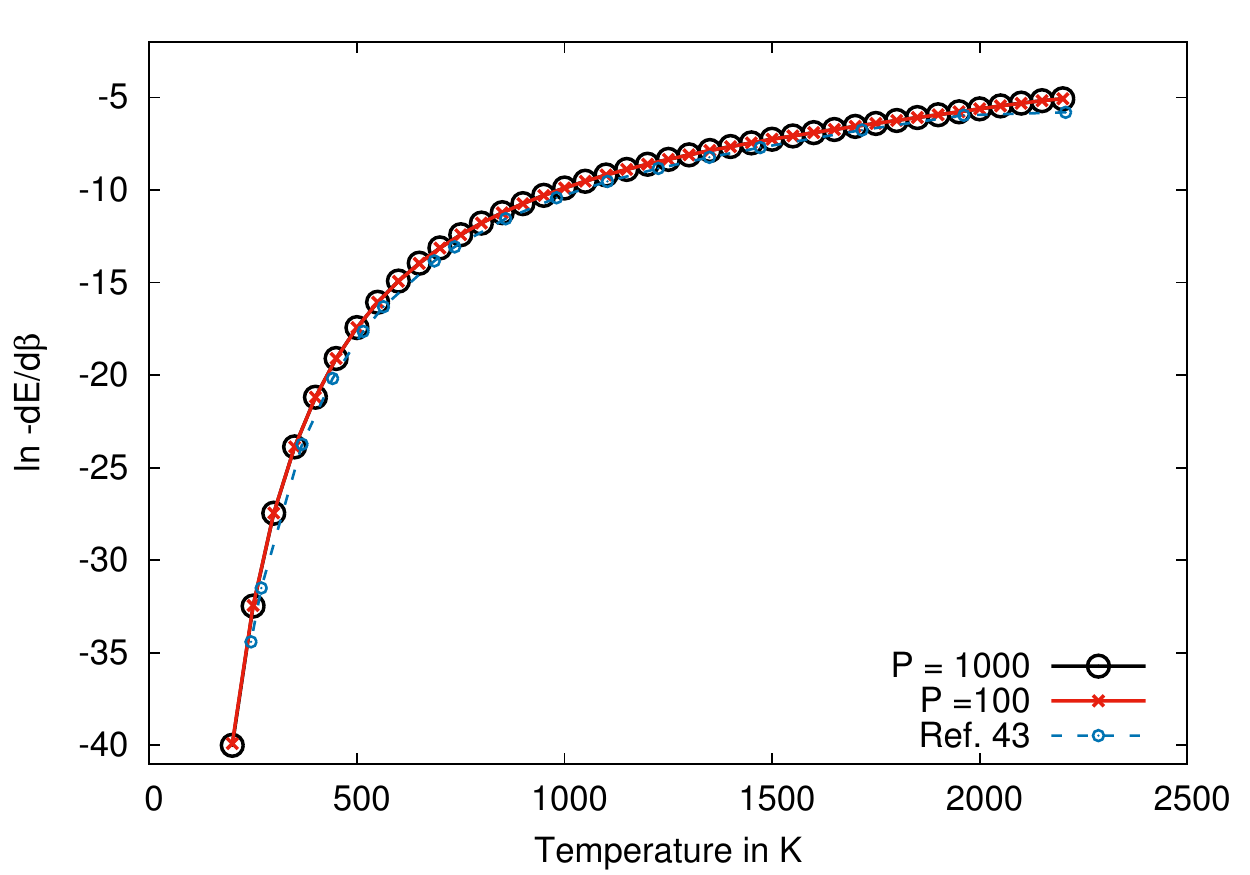} 
\caption{Change of the instanton's tunneling energy with respect to $\beta$
  in the M\"uller--Brown potential calculated at different temperatures for a
  different number of images $P$. The value of $\frac{dE}{d\beta}$ is negative
  for all $T < T_c$. The dashed blue line uses the approximation scheme previously
  proposed in Ref. \citenum{mcconnel171}.}
\label{pic111}
\end{figure}

\begin{figure}
\includegraphics[width=8cm]{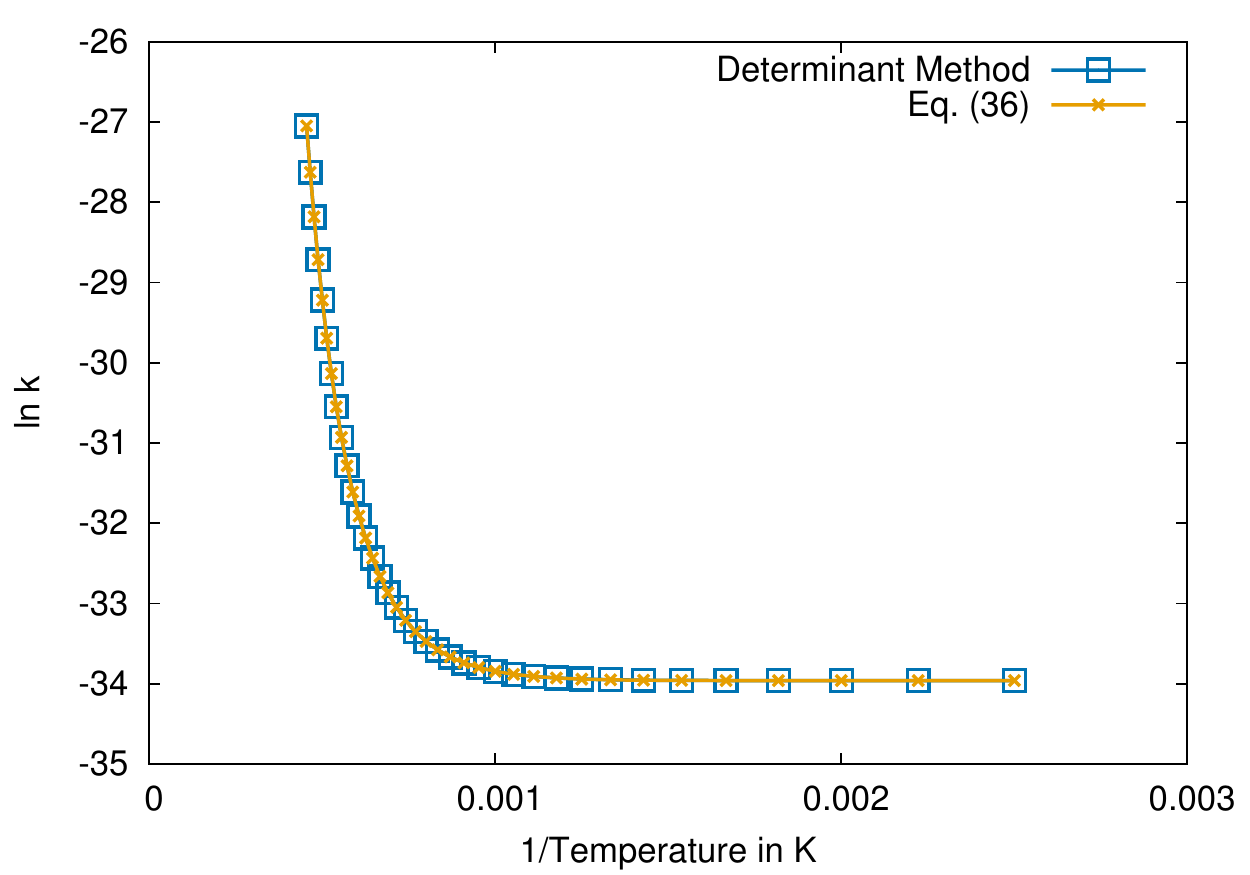}
\caption{Rate constant in atomic units for the M\"uller--Brown potential. The orange
  line was calculated using \eqref{final_final} and the quantities determined
  in \figref{pic11} and \figref{pic111}. The blue line was calculated with
  the conventional determinant method.}
\label{pic12}
\end{figure}

In the M\"uller--Brown potential\cite{mul79} we study the transition from the
intermediate minimum at the coordinates $\left( -0.05001, 0.46669 \right)$
over the barrier with the saddle point at $\left( -0.822001, 0.624314 \right)$
to the global minimum. Since this is a two-dimensional system without
rotational and translational invariance, one stability parameter is zero and
another one represents the fluctuations orthogonal to the instanton path. For a
particle of the mass of a hydrogen atom moving in the M\"uller--Brown
potential, the crossover temperature is 2207~K. Figs.~\ref{pic11},
\ref{pic111} and \ref{pic12} show the results for the non zero stability
parameter $u$, the quantity $\frac{dE}{d\beta}$ and consequently the thermal
rate constant $k(\beta)$, which can directly be calculated from these
quantities using \eqref{final_final}. 

\Figref{pic11} demonstrates that the value of the stability parameter is
rather independent of the number of images. Only at rather low temperature
compared to $T_\text{c}$, where the rate constant is already pretty
independent of the temperature are more images required to achieve a converged
result for $u$. The largest deviation occurs at the lowest temperature, $T =
350$~K, with a deviation of only 1.3\% between $P=100$ and
$P=1000$. \Figref{pic111} compares the result of the calculation for
$\frac{dE}{d\beta}$ using the solution of \eqref{5.5} to an approximation
scheme previously proposed in Ref. \citenum{mcconnel171}. They all agree very
well.

If we compare the results for $k(T)$ in \figref{pic12} calculated with the
conventional determinant method (blue) and the new approach (orange), both
with 1000 images, we get to almost identical results. The largest deviation
between the two methods was 0.06\% at $T = 450$~K long after the instanton
converges to the final rate constant.

\subsection{DFT Calculations for the Methylhydroxycarbene}

The tunneling-decay of methylhydroxycarbene to acetaldehyde was
previously studied experimentally and
computationally.\cite{sch11a,kae13} Here, we use this reaction to
investigate the stability of our new method with respect to numerical
noise in the gradients and Hessians and to provide data for a system
with translational and rotational invariance. We obtained energies and
their derivatives on-the-fly using DFT calculations with the B3LYP
hybrid functional \cite{dir29,sla51,vos80,bec88,lee88,bec93} in
combination with the def2-SVP basis set.\cite{weigend2005} These
calculations were done using Turbomole v 7.0.1.\cite{turbomole} For
the geometry optimizations and and subsequent instanton calculations
DL-FIND \cite{kae09a} was used via ChemShell \cite{met14} as interface
to Turbomole. Note that here we compare methods rather than aiming at
high accuracy in comparison to experiment. Nevertheless, B3LYP (with a
different basis set, though) was found to reproduce more accurate
calculations quite well.\cite{kae13}

In the calculations
$P=128$ images were used for the discretization of the instanton path
unless notes otherwise.
Instantons have been optimized such that the largest component of the
gradient is smaller than $10^{-9}$ atomic units. The SCF cycles were
iterated up to the change was less than $10^{-9}$ Hartree per
iteration. The $m5$ grid\cite{eichkorn1997} was used.

\begin{figure}
\includegraphics[width=8cm]{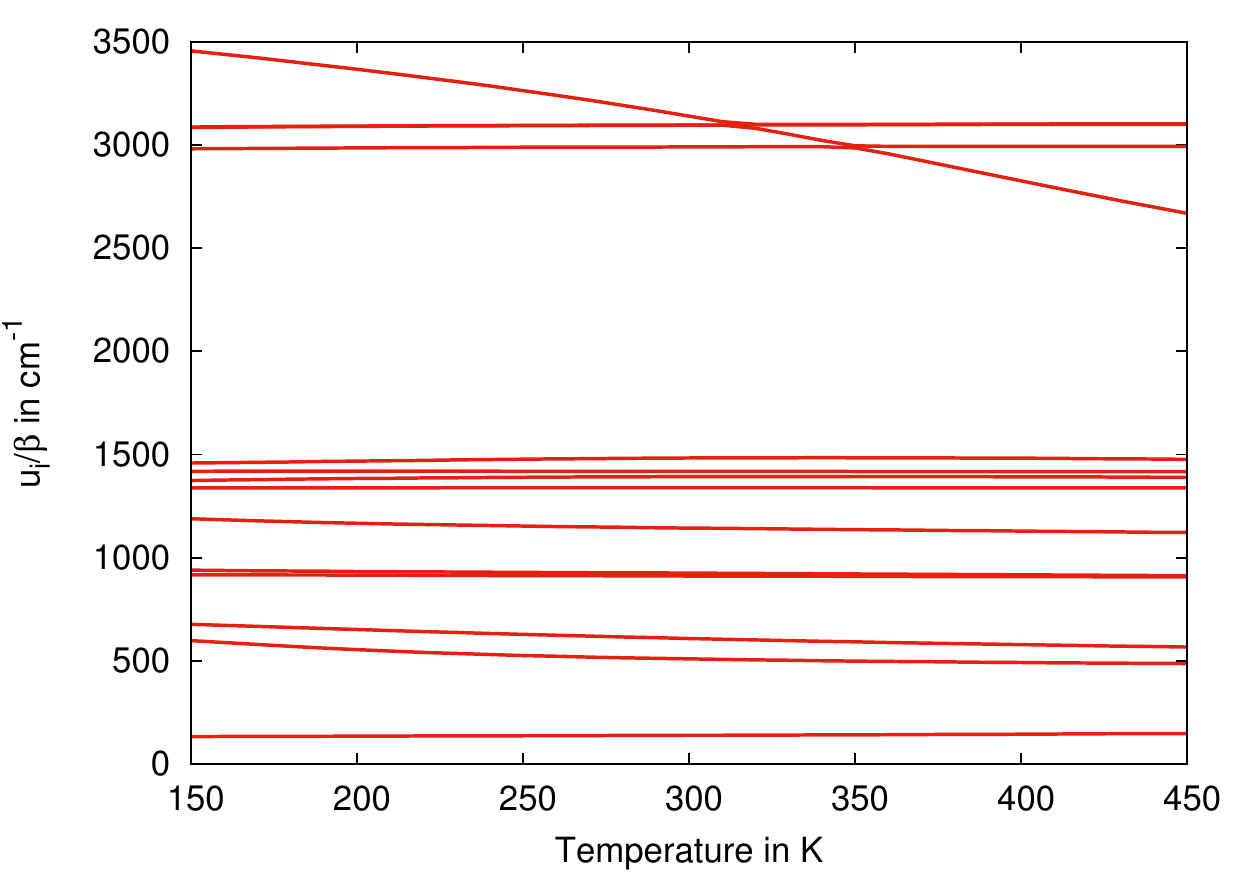}% \\
\caption{The 14 non-zero stability parameters for the carbene system using
  128 images.% Bottom: the natural logarithm of $F_\text{Inst} =
  %\prod_{i=1}^{14} \left( 2 \sinh( \frac{1}{2} u_i)\right)^{-1} $.
}
\label{pic13}
\end{figure}

% TS: 2660.665 cm^-1 ist combined symmetric C-H O-H stretch mode (transferred H)
% Min (Carbene) 3735.989  cm^-1 ist OH-stretch.

 After optimizing the instanton geometries at a series of temperatures
below the crossover temperature of 453~K, we calculated the
stability parameters $u_i$ using \eqref{foh4}. For this 7-atom system
there are 14 non-zero stability parameters. They can be interpreted
as frequencies via $\omega_i=u_i/\beta$ as discussed
previously.\cite{mcconnel17} These stability parameters are depicted
in \figref{pic13}. Most of them are almost independent of $T$. The
stability parameter with the strongest $T$-dependence corresponds to
the movement of the transferred H-atom perpendicular to the instanton
path. At high temperature, the instanton path is short and in direct
vicinity of the saddle point. Correspondingly, the value of the
$T$-dependent stability parameter is close to  2660.7~cm$^{-1}$, the
wave number of the C--H and O--H stretching mode of the transferred
H. At lower $T$ its value increases, tending towards the value of
3736.0~cm$^{-1}$, which corresponds to the O--H stretching mode in the
reactant, methylhydroxycarbene. %The resulting fluctuation factor
%$F_\text{Inst}$ is shown in the lower part of \figref{pic13}. 

\begin{figure}
\includegraphics[width=8cm]{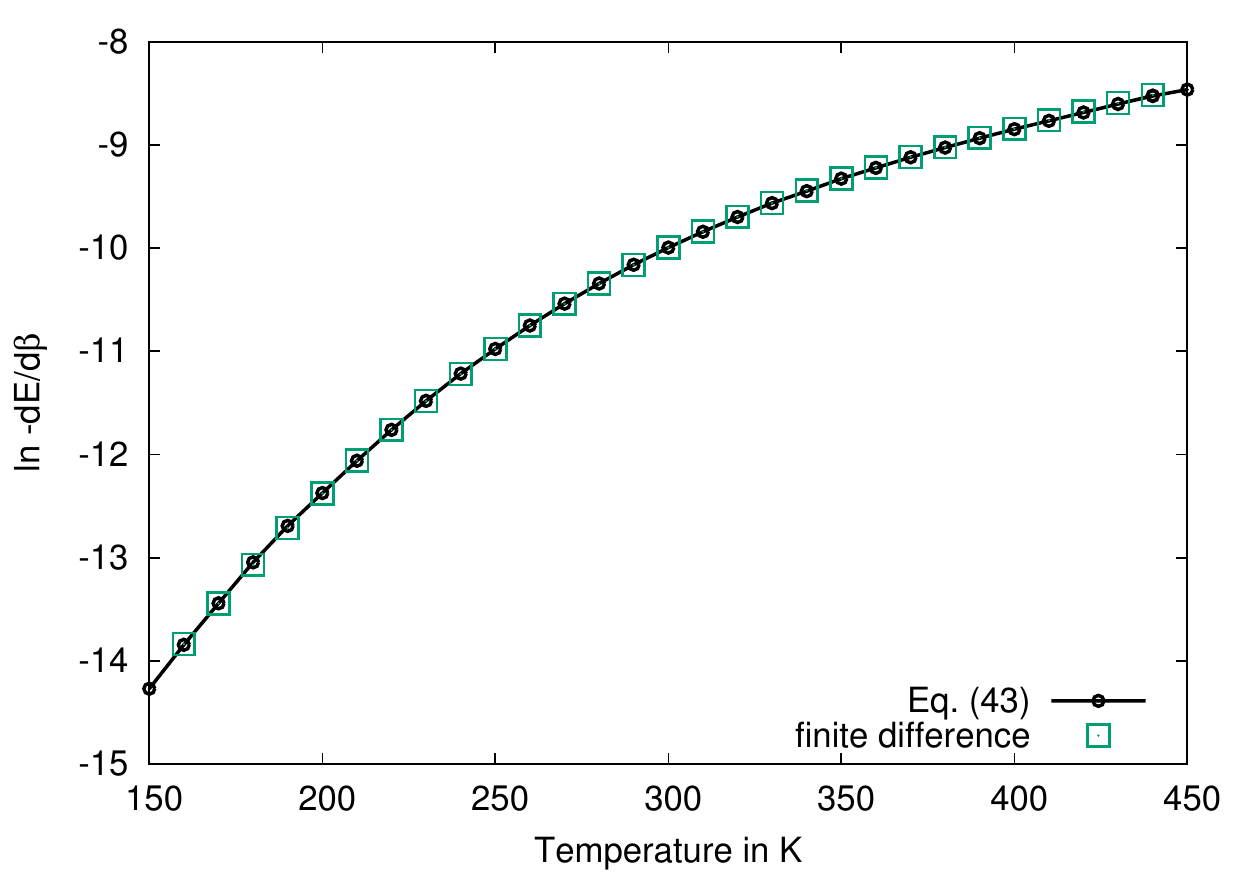} 
\caption{Change of the instanton's tunneling energy with respect to $\beta$ in
  for the reaction of methylhydroxycarbene to acetaldehyde calculated with 128
  images. The value of $\frac{dE}{d\beta}$ is negative for all $T < T_c$. The
  green squares correspond to finite differences between $E$ obtained for
  instantons at adjacent temperatures. }
\label{picdedbc}
\end{figure}

The temperature-dependence of $dE/d\beta$ is shown in \figref{picdedbc}. An
approximative previous approach is shown for comparison. Our new
approach agrees very well with a finite-difference estimation from
instantons at different temperatures. 

\begin{figure}
\includegraphics[width=8cm]{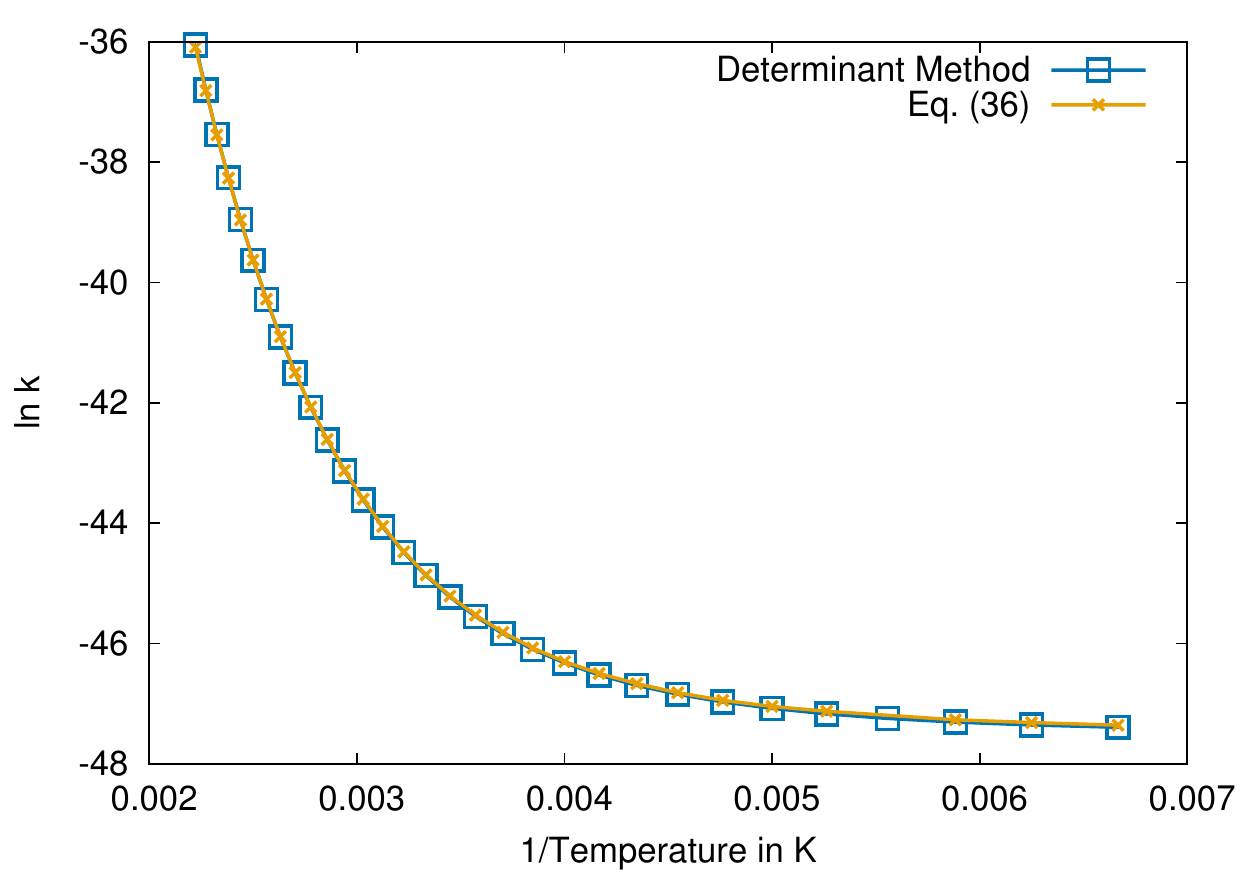}
\caption{Rate constants in atomic units for the reaction of
  methylhydroxycarbene to acetaldehyde. The orange line was calculated
  using \eqref{eq10} and the quantities determined in
  \figref{pic13}. The blue line was calculated with the conventional
  determinant method.}
\label{pic14}
\end{figure}

 Using the results in \figref{pic13} and \figref{picdedbc} and the
classical expressions for $Q_\text{t-r}$ we obtain the
thermal rate constants depicted in \figref{pic14}. One can see clearly a very
good agreement between the new approach and the conventional determinant
method. The highest deviations between the methods appear at low
temperature with the rate constant determined by the
new approach being 4.3\% larger than the one calculated via the determinant
method. % it is at 190K

\begin{figure}
\includegraphics[width=8cm]{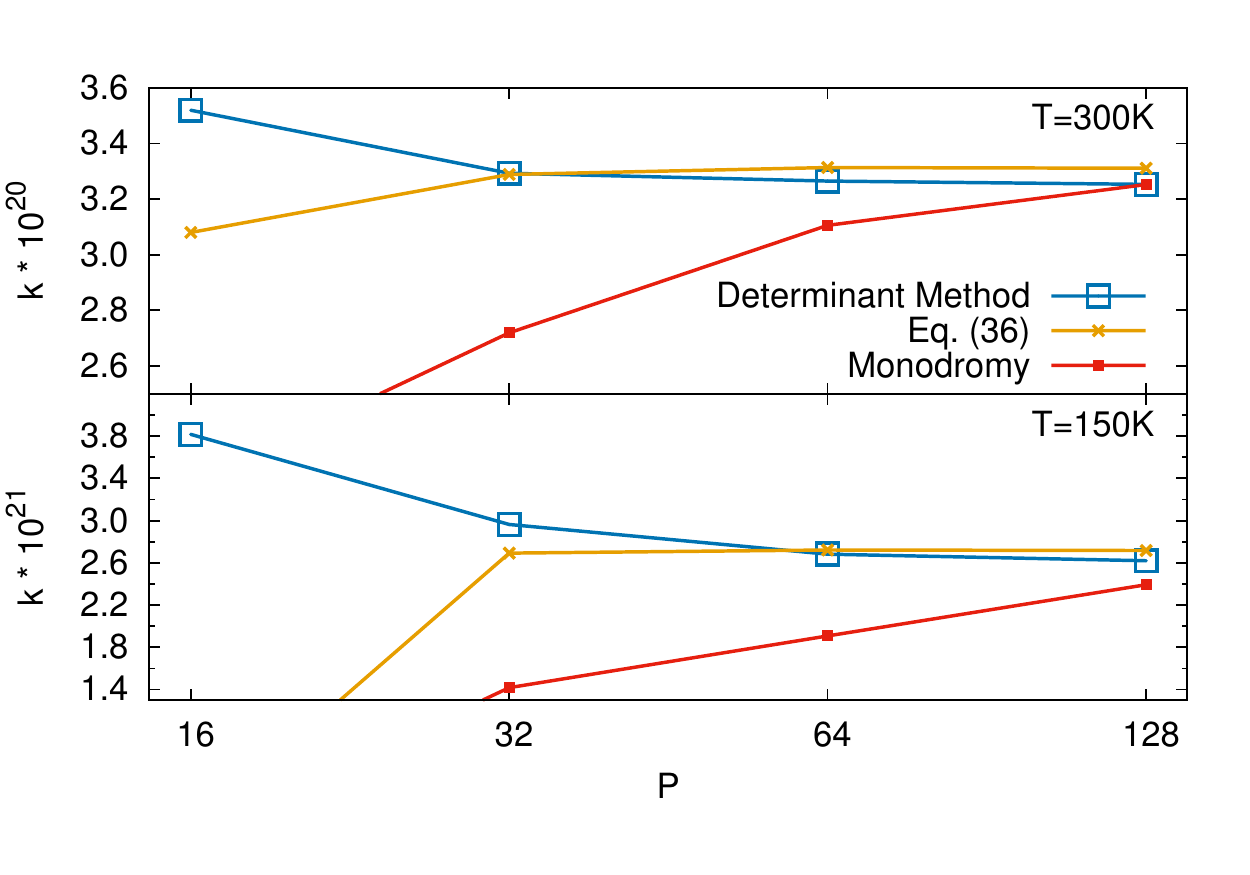}
\caption{Convergence of the rate constants with the number of images $P$ for
  two temperatures and for the reaction of methylhydroxycarbene to
  acetaldehyde. The orange line was calculated using \eqref{eq10} and the
  quantities determined in \figref{pic13}. The blue line was calculated with
  the conventional determinant method. \new{The red line was obtained from
    integrating \eqref{mono} with $dE/d\beta$ from \eqref{eq9}.}
\label{pic8}}
\end{figure}

The convergence of both methods with $P$ is displayed in \figref{pic8} for two
temperatures, $T=300$~K (top) and $T=150$~K (bottom) \new{and compared to the
  standard approach of calculating the stability parameters $u_i$ from the
  monodromy matrix obtained by integrating \eqref{mono}.\cite{mcconnel17}} For
very few images, \new{all three} methods are inaccurate, but the determinant
method is slightly more stable than our new approach. An instanton path with
merely 16 images at a temperature so much below $T_\text{c}$ results in a
rather badly discretized path. At higher numbers of images, it is notable that
the new method converges faster to the final result than the determinant
method. \new{The integration of \eqref{mono} is found to converge slowest with
$P$.} This is a trend we have observed in general. It is most clearly visible
at $T=150$~K.  Note the different scale of the vertical axes for the two
graphs. A higher number of images is computationally easier accessible with
the new method, because its computational effort scales lower with the number
of images.

%rates_image_comparision.txt     dlf 64  Andi 64
%rate_carbene.txt   dlf 64  4:andi64

%%%%%%%%%%%%%%%%%%%%%%%%%%%%%%%%%%%%%%%%%%%%%%%%%%%%%%%%%%%%%%%%%%%%%%%%

\section{Conclusion}

When we compare the results of our new algorithm with the determinant method
we obtain practically the same rate constants, for the analytical as well as the DFT
potential, yet the computational effort for the rate constant calculation
is reduced as no matrix diagonalization is required. Instead of
diagonalizing a $PD \times PD$ matrix one has to solve two different linear
systems of equations. % if \eqref{final_final} is used instead of
%\eqref{soso}. 
For the calculation of $\frac{dE}{d\beta}$ one has to solve the system in
\eqref{5.5}, which requires Hessians as well as gradients at each point of the
trajectory. Gradients are not required in the determinant method. However,
they are required for the instanton optimization, so no additional effort is
needed. The calculation of the fluctuation factor of the instanton is done by
solving \eqref{foh4} for two different right-hand sides. Furthermore the
proposed method improves on the previously presented
approaches\cite{mcconnel17} by providing a rigorously derived and numerically
stable way of obtaining stability parameters in multidimensional
systems. These are important in the context of microcanonical and canonical
instanton theory. While the fundamental bottleneck of any reaction rate
constant calculation remains the calculation of energies and Hessians, the new
approach might be particularly helpful if one is interested in obtaining
thermal rate constants at very low temperature for a system for which a fitted
analytic PES is available and therefore a high number of images can be
computed. In these cases the rate constant calculation can be significantly
speeded up.
	
%%%%%%%%%%%%%%%%%%%%%%%%%%%%%%%%%%%%%%%%%%%%%%%%%%%%%%%%%%%%%%%%%%%%%%%% 
%\begin{acknowledgments} 

\section*{Acknowledgments}
The authors thank Dr. J. Meisner for performing the
  DFT calculations. This work was financially supported by the European
  Union's Horizon 2020 research and innovation programme (grant agreement
  No. 646717, TUNNELCHEM). A.L. is financially supported by the Carl-Zeiss
  Foundation.

    \section{Appendix}
     \subsection{Analytic Calculation of $Q_\text{RS}$}
    \label{Calc_QRS}
The trajectory of  $D$ uncoupled harmonic oscillators in imaginary time is given by
\begin{align}
\mathbf{X}(\tau) = \frac{ x_i'' \sinh(\omega_i \tau) - x_i'' \sinh((\tau - \beta)\omega_i) }{\sinh(\beta \omega_i)} \  \hat{\mathbf{e}}_i 
\end{align}
for $  i \in \left[ 1, \dots, D \right] $. The corresponding Euclidean action is
\begin{align}
\mathcal{S}_\text{E}(\mathbf{x}', \mathbf{x}'', \beta)  &= \int_0^{\beta\hbar} \left( \frac{1}{2} \dot{\mathbf{X}}^2 + \frac{1}{2} \sum_i^D \omega_i^2 x_i^2(\tau) \right)\ d\tau \\
&= \sum_{i=1}^D \frac{ \left( x_i'^2 + x_i''^2 \right) \cosh(\omega_i \beta) - 2 x_i'' x_i' }{\sinh(\omega_i \beta)}	.
\end{align}
The evaluation of the second derivatives of $\mathcal{S}_\text{E}$ evaluated at the point $\mathbf{x}'' = \mathbf{x}' = \mathbf{x}$ yields
\begin{align} 
\frac{\partial^2 \mathcal{S}_\text{E}}{\partial \mathbf{x}'_i \partial \mathbf{x}''_j }\Bigr|_{\substack{x' = x''}} &= -\frac{\omega_i}{\sinh(\omega_i \beta)} \delta_{i,j} \\  
\frac{\partial^2 \mathcal{S}_\text{E}}{\partial \mathbf{x}'_i \partial \mathbf{x}''_j }\Bigr|_{\substack{x' = x''}} &= \omega_i \coth(\omega_i \beta) \delta_{i,j}  \\ 
\frac{\partial^2 \mathcal{S}_\text{E}}{\partial \mathbf{x}'_i \partial \mathbf{x}''_j }\Bigr|_{\substack{x' = x''}} &= \omega_i \coth(\omega_i \beta) \delta_{i,j} .
\end{align}
Using \eqref{mm11} then yields
\begin{align}
F &= \sqrt{\frac{ \left| - \mathbf{b}\right|}{\left| \mathbf{a} + 2 \mathbf{b} + \mathbf{c}\right|}} = \sqrt{\frac{(-1)^D}{\left| \mathbf{M} - \mathbf{1} \right|}} \\
&= \prod_{i=1}^D \frac{1}{2 \sinh(\frac{\omega_i}{2} \beta)}
\end{align}
which gives for the $\mathcal{Q}_\text{RS}$
\begin{align}
\mathcal{Q}_\text{RS} &= \prod_{i=1}^D \frac{1}{2 \sinh(\frac{\omega_i}{2} \beta)} \ee^{-\beta V(\mathbf{x}_\text{RS})},
\end{align}
i.e. the analytic solution of the partition function of a multidimensional
harmonic oscillator.
%which gives
%\begin{align}
%\mathcal{Q}_\text{RS} = \prod_{i=1}^D \frac{1}{2 \sinh(\frac{\omega_i}{2} \beta)} 
%\end{align}
%as the energy can be shifted by an arbitrary constant without loss of generality such that %$V(\mathbf{x}_\text{RS}) = 0$.

\bibliography{jabref}

\providecommand{\latin}[1]{#1}
\providecommand*\mcitethebibliography{\thebibliography}
\csname @ifundefined\endcsname{endmcitethebibliography}
  {\let\endmcitethebibliography\endthebibliography}{}
\begin{mcitethebibliography}{62}
\providecommand*\natexlab[1]{#1}
\providecommand*\mciteSetBstSublistMode[1]{}
\providecommand*\mciteSetBstMaxWidthForm[2]{}
\providecommand*\mciteBstWouldAddEndPuncttrue
  {\def\EndOfBibitem{\unskip.}}
\providecommand*\mciteBstWouldAddEndPunctfalse
  {\let\EndOfBibitem\relax}
\providecommand*\mciteSetBstMidEndSepPunct[3]{}
\providecommand*\mciteSetBstSublistLabelBeginEnd[3]{}
\providecommand*\EndOfBibitem{}
\mciteSetBstSublistMode{f}
\mciteSetBstMaxWidthForm{subitem}{(\alph{mcitesubitemcount})}
\mciteSetBstSublistLabelBeginEnd
  {\mcitemaxwidthsubitemform\space}
  {\relax}
  {\relax}

\bibitem[Miyazaki(2004)]{miy04}
Miyazaki,~T., Ed. \emph{Atom Tunneling Phenomena in Physics, Chemistry and
  Biology}; Springer, Berlin, Germany, 2004\relax
\mciteBstWouldAddEndPuncttrue
\mciteSetBstMidEndSepPunct{\mcitedefaultmidpunct}
{\mcitedefaultendpunct}{\mcitedefaultseppunct}\relax
\EndOfBibitem
\bibitem[Kohen and Limbach(2005)Kohen, and Limbach]{koh05}
Kohen,~A., Limbach,~H.-H., Eds. \emph{Isotope Effects in Chemistry and
  Biology}; CRC Press, Boca Raton, FL, USA, 2005\relax
\mciteBstWouldAddEndPuncttrue
\mciteSetBstMidEndSepPunct{\mcitedefaultmidpunct}
{\mcitedefaultendpunct}{\mcitedefaultseppunct}\relax
\EndOfBibitem
\bibitem[Allemann and Scrutton(2009)Allemann, and Scrutton]{all09}
Allemann,~R.~K., Scrutton,~N.~S., Eds. \emph{Quantum Tunnelling in
  Enzyme-Catalysed Reactions}; RSC Publishing, Cambridge, UK, 2009\relax
\mciteBstWouldAddEndPuncttrue
\mciteSetBstMidEndSepPunct{\mcitedefaultmidpunct}
{\mcitedefaultendpunct}{\mcitedefaultseppunct}\relax
\EndOfBibitem
\bibitem[Kohen(2003)]{koh03}
Kohen,~A. Kinetic Isotope Effects as Probes for Hydrogen Tunneling, Coupled
  Motion and Dynamics Contributions to Enzyme Catalysis. \emph{Prog. React.
  Kinet. Mech.} \textbf{2003}, \emph{28}, 119--156\relax
\mciteBstWouldAddEndPuncttrue
\mciteSetBstMidEndSepPunct{\mcitedefaultmidpunct}
{\mcitedefaultendpunct}{\mcitedefaultseppunct}\relax
\EndOfBibitem
\bibitem[Nagel and Klinman(2006)Nagel, and Klinman]{nag06}
Nagel,~Z.~D.; Klinman,~J.~P. Tunneling and Dynamics in Enzymatic Hydride
  Transfer. \emph{Chem. Rev.} \textbf{2006}, \emph{106}, 3095--3118\relax
\mciteBstWouldAddEndPuncttrue
\mciteSetBstMidEndSepPunct{\mcitedefaultmidpunct}
{\mcitedefaultendpunct}{\mcitedefaultseppunct}\relax
\EndOfBibitem
\bibitem[Borden(2016)]{bor16}
Borden,~W.~T. Reactions that involve tunneling by carbon and the role that
  calculations have played in their study. \emph{WIREs Comput. Mol. Sci.}
  \textbf{2016}, \emph{6}, 20--46\relax
\mciteBstWouldAddEndPuncttrue
\mciteSetBstMidEndSepPunct{\mcitedefaultmidpunct}
{\mcitedefaultendpunct}{\mcitedefaultseppunct}\relax
\EndOfBibitem
\bibitem[Meisner and K\"astner(2016)Meisner, and K\"astner]{mei16}
Meisner,~J.; K\"astner,~J. Atom-Tunneling in Chemistry. \emph{Angew. Chem. Int.
  Ed.} \textbf{2016}, \emph{55}, 5400--5413\relax
\mciteBstWouldAddEndPuncttrue
\mciteSetBstMidEndSepPunct{\mcitedefaultmidpunct}
{\mcitedefaultendpunct}{\mcitedefaultseppunct}\relax
\EndOfBibitem
\bibitem[Fern\'andez-Ramos \latin{et~al.}(2006)Fern\'andez-Ramos, Miller,
  Klippenstein, and Truhlar]{fer06}
Fern\'andez-Ramos,~A.; Miller,~J.~A.; Klippenstein,~S.~J.; Truhlar,~D.~G.
  Modeling the Kinetics of Bimolecular Reactions. \emph{Chem. Rev.}
  \textbf{2006}, \emph{106}, 4518--4584\relax
\mciteBstWouldAddEndPuncttrue
\mciteSetBstMidEndSepPunct{\mcitedefaultmidpunct}
{\mcitedefaultendpunct}{\mcitedefaultseppunct}\relax
\EndOfBibitem
\bibitem[Pu \latin{et~al.}(2006)Pu, Gao, and Truhlar]{pu06}
Pu,~J.; Gao,~J.; Truhlar,~D.~G. Multidimensional Tunneling, Recrossing, and the
  Transmission Coefficient for Enzymatic Reactions. \emph{Chem. Rev.}
  \textbf{2006}, \emph{106}, 3140--3169\relax
\mciteBstWouldAddEndPuncttrue
\mciteSetBstMidEndSepPunct{\mcitedefaultmidpunct}
{\mcitedefaultendpunct}{\mcitedefaultseppunct}\relax
\EndOfBibitem
\bibitem[Nyman(2014)]{nym14}
Nyman,~G. Computational methods of quantum reaction dynamics. \emph{Int. J.
  Quant. Chem.} \textbf{2014}, \emph{114}, 1183--1198\relax
\mciteBstWouldAddEndPuncttrue
\mciteSetBstMidEndSepPunct{\mcitedefaultmidpunct}
{\mcitedefaultendpunct}{\mcitedefaultseppunct}\relax
\EndOfBibitem
\bibitem[K\"astner(2014)]{kae14}
K\"astner,~J. Theory and Simulation of Atom Tunneling in Chemical Reactions.
  \emph{WIREs Comput. Mol. Sci.} \textbf{2014}, \emph{4}, 158--168\relax
\mciteBstWouldAddEndPuncttrue
\mciteSetBstMidEndSepPunct{\mcitedefaultmidpunct}
{\mcitedefaultendpunct}{\mcitedefaultseppunct}\relax
\EndOfBibitem
\bibitem[Langer(1967)]{lan67}
Langer,~J.~S. Theory of the condensation point. \emph{Ann. Phys. (N.Y.)}
  \textbf{1967}, \emph{41}, 108--157\relax
\mciteBstWouldAddEndPuncttrue
\mciteSetBstMidEndSepPunct{\mcitedefaultmidpunct}
{\mcitedefaultendpunct}{\mcitedefaultseppunct}\relax
\EndOfBibitem
\bibitem[Langer(1969)]{lan69}
Langer,~J.~S. Statistical theory of the decay of metastable states. \emph{Ann.
  Phys. (N.Y.)} \textbf{1969}, \emph{54}, 258--275\relax
\mciteBstWouldAddEndPuncttrue
\mciteSetBstMidEndSepPunct{\mcitedefaultmidpunct}
{\mcitedefaultendpunct}{\mcitedefaultseppunct}\relax
\EndOfBibitem
\bibitem[Miller(1975)]{mil75}
Miller,~W.~H. Semiclassical limit of quantum mechanical transition state theory
  for nonseparable systems. \emph{J. Chem. Phys.} \textbf{1975}, \emph{62},
  1899--1906\relax
\mciteBstWouldAddEndPuncttrue
\mciteSetBstMidEndSepPunct{\mcitedefaultmidpunct}
{\mcitedefaultendpunct}{\mcitedefaultseppunct}\relax
\EndOfBibitem
\bibitem[Coleman(1977)]{col77}
Coleman,~S. Fate of the false vacuum: Semiclassical theory. \emph{Phys. Rev. D}
  \textbf{1977}, \emph{15}, 2929--2936\relax
\mciteBstWouldAddEndPuncttrue
\mciteSetBstMidEndSepPunct{\mcitedefaultmidpunct}
{\mcitedefaultendpunct}{\mcitedefaultseppunct}\relax
\EndOfBibitem
\bibitem[Callan~Jr. and Coleman(1977)Callan~Jr., and Coleman]{cal77}
Callan~Jr.,~C.~G.; Coleman,~S. Fate of the false vacuum. II. First quantum
  corrections. \emph{Phys. Rev. D} \textbf{1977}, \emph{16}, 1762--1768\relax
\mciteBstWouldAddEndPuncttrue
\mciteSetBstMidEndSepPunct{\mcitedefaultmidpunct}
{\mcitedefaultendpunct}{\mcitedefaultseppunct}\relax
\EndOfBibitem
\bibitem[Affleck(1981)]{aff81}
Affleck,~I. Quantum-Statistical Metastability. \emph{Phys. Rev. Lett.}
  \textbf{1981}, \emph{46}, 388--391\relax
\mciteBstWouldAddEndPuncttrue
\mciteSetBstMidEndSepPunct{\mcitedefaultmidpunct}
{\mcitedefaultendpunct}{\mcitedefaultseppunct}\relax
\EndOfBibitem
\bibitem[Benderskii \latin{et~al.}(1994)Benderskii, Makarov, and Wight]{ben94}
Benderskii,~V.~A.; Makarov,~D.~E.; Wight,~C.~A. One-Dimensional Models.
  \emph{Adv. Chem. Phys.} \textbf{1994}, \emph{88}, 55--95\relax
\mciteBstWouldAddEndPuncttrue
\mciteSetBstMidEndSepPunct{\mcitedefaultmidpunct}
{\mcitedefaultendpunct}{\mcitedefaultseppunct}\relax
\EndOfBibitem
\bibitem[Feynman and Hibbs(1965)Feynman, and Hibbs]{Feynman_book}
Feynman,~R.~P.; Hibbs,~A.~R. \emph{Quantum Mechanics and Path Integrals};
  McGraw-Hill, 1965\relax
\mciteBstWouldAddEndPuncttrue
\mciteSetBstMidEndSepPunct{\mcitedefaultmidpunct}
{\mcitedefaultendpunct}{\mcitedefaultseppunct}\relax
\EndOfBibitem
\bibitem[Arnaldsson(2007)]{arn07}
Arnaldsson,~A. Calculation of quantum mechanical rate constants directly from
  ab initio atomic forces. Ph.D.\ thesis, University of Washington, 2007\relax
\mciteBstWouldAddEndPuncttrue
\mciteSetBstMidEndSepPunct{\mcitedefaultmidpunct}
{\mcitedefaultendpunct}{\mcitedefaultseppunct}\relax
\EndOfBibitem
\bibitem[Andersson \latin{et~al.}(2009)Andersson, Nyman, Arnaldsson, Manthe,
  and J{\'o}nsson]{and09}
Andersson,~S.; Nyman,~G.; Arnaldsson,~A.; Manthe,~U.; J{\'o}nsson,~H.
  Comparison of Quantum Dynamics and Quantum Transition State Theory Estimates
  of the {H + CH$_4$} Reaction Rate. \emph{J. Phys. Chem. A} \textbf{2009},
  \emph{113}, 4468--4478\relax
\mciteBstWouldAddEndPuncttrue
\mciteSetBstMidEndSepPunct{\mcitedefaultmidpunct}
{\mcitedefaultendpunct}{\mcitedefaultseppunct}\relax
\EndOfBibitem
\bibitem[Rommel \latin{et~al.}(2011)Rommel, Goumans, and K\"astner]{rom11}
Rommel,~J.~B.; Goumans,~T. P.~M.; K\"astner,~J. Locating instantons in many
  degrees of freedom. \emph{J. Chem. Theory Comput.} \textbf{2011}, \emph{7},
  690--698\relax
\mciteBstWouldAddEndPuncttrue
\mciteSetBstMidEndSepPunct{\mcitedefaultmidpunct}
{\mcitedefaultendpunct}{\mcitedefaultseppunct}\relax
\EndOfBibitem
\bibitem[Rommel and K\"astner(2011)Rommel, and K\"astner]{rom11b}
Rommel,~J.~B.; K\"astner,~J. Adaptive Integration Grids in Instanton Theory
  Improve the Numerical Accuracy at Low Temperature. \emph{J. Chem. Phys.}
  \textbf{2011}, \emph{134}, 184107\relax
\mciteBstWouldAddEndPuncttrue
\mciteSetBstMidEndSepPunct{\mcitedefaultmidpunct}
{\mcitedefaultendpunct}{\mcitedefaultseppunct}\relax
\EndOfBibitem
\bibitem[Faddeev and Popov(1967)Faddeev, and Popov]{FADDEEV67}
Faddeev,~L.; Popov,~V. Feynman diagrams for the Yang--Mills field. \emph{Phys.
  Lett. B} \textbf{1967}, \emph{25}, 29--30\relax
\mciteBstWouldAddEndPuncttrue
\mciteSetBstMidEndSepPunct{\mcitedefaultmidpunct}
{\mcitedefaultendpunct}{\mcitedefaultseppunct}\relax
\EndOfBibitem
\bibitem[Kleinert(2009)]{kle09}
Kleinert,~H. \emph{Path Integrals in Quantum Mechanics, Statistics, Polymer
  Physics, and Financial Markets}, 5th ed.; World Scientific, 2009\relax
\mciteBstWouldAddEndPuncttrue
\mciteSetBstMidEndSepPunct{\mcitedefaultmidpunct}
{\mcitedefaultendpunct}{\mcitedefaultseppunct}\relax
\EndOfBibitem
\bibitem[Coleman(1988)]{col88}
Coleman,~S. Quantum Tunneling and negative Eigenvalues. \emph{Nucl. Phys. B}
  \textbf{1988}, \emph{298}, 178--186\relax
\mciteBstWouldAddEndPuncttrue
\mciteSetBstMidEndSepPunct{\mcitedefaultmidpunct}
{\mcitedefaultendpunct}{\mcitedefaultseppunct}\relax
\EndOfBibitem
\bibitem[H\"anggi \latin{et~al.}(1990)H\"anggi, Talkner, and Borkovec]{han90}
H\"anggi,~P.; Talkner,~P.; Borkovec,~M. Reaction-rate theory: fifty years after
  Kramers. \emph{Rev. Mod. Phys.} \textbf{1990}, \emph{62}, 251--341\relax
\mciteBstWouldAddEndPuncttrue
\mciteSetBstMidEndSepPunct{\mcitedefaultmidpunct}
{\mcitedefaultendpunct}{\mcitedefaultseppunct}\relax
\EndOfBibitem
\bibitem[Messina \latin{et~al.}(1995)Messina, Schenter, and Garrett]{mes95}
Messina,~M.; Schenter,~G.~K.; Garrett,~B.~C. A variational centroid density
  procedure for the calculation of transmission coefficients for asymmetric
  barriers at low temperature. \emph{J. Chem. Phys.} \textbf{1995}, \emph{103},
  3430\relax
\mciteBstWouldAddEndPuncttrue
\mciteSetBstMidEndSepPunct{\mcitedefaultmidpunct}
{\mcitedefaultendpunct}{\mcitedefaultseppunct}\relax
\EndOfBibitem
\bibitem[Richardson and Althorpe(2009)Richardson, and Althorpe]{ric09}
Richardson,~J.~O.; Althorpe,~S.~C. Ring-polymer molecular dynamics rate-theory
  in the deep-tunneling regime: Connection with semiclassical instanton theory.
  \emph{J. Chem. Phys.} \textbf{2009}, \emph{131}, 214106\relax
\mciteBstWouldAddEndPuncttrue
\mciteSetBstMidEndSepPunct{\mcitedefaultmidpunct}
{\mcitedefaultendpunct}{\mcitedefaultseppunct}\relax
\EndOfBibitem
\bibitem[Kryvohuz(2011)]{kry11}
Kryvohuz,~M. Semiclassical instanton approach to calculation of reaction rate
  constants in multidimensional chemical systems. \emph{J. Chem. Phys.}
  \textbf{2011}, \emph{134}, 114103\relax
\mciteBstWouldAddEndPuncttrue
\mciteSetBstMidEndSepPunct{\mcitedefaultmidpunct}
{\mcitedefaultendpunct}{\mcitedefaultseppunct}\relax
\EndOfBibitem
\bibitem[Althorpe(2011)]{alt11}
Althorpe,~S.~C. On the equivalence of two commonly used forms of semiclassical
  instanton theory. \emph{J. Chem. Phys.} \textbf{2011}, \emph{134},
  114104\relax
\mciteBstWouldAddEndPuncttrue
\mciteSetBstMidEndSepPunct{\mcitedefaultmidpunct}
{\mcitedefaultendpunct}{\mcitedefaultseppunct}\relax
\EndOfBibitem
\bibitem[Richardson(2018)]{ric18}
Richardson,~J.~O. Ring-polymer instanton theory. \emph{J. Chem. Phys.}
  \textbf{2018}, \emph{148}, 200901\relax
\mciteBstWouldAddEndPuncttrue
\mciteSetBstMidEndSepPunct{\mcitedefaultmidpunct}
{\mcitedefaultendpunct}{\mcitedefaultseppunct}\relax
\EndOfBibitem
\bibitem[Richardson(2016)]{ric16}
Richardson,~J.~O. Derivation of instanton rate theory from first principles.
  \emph{J. Chem. Phys.} \textbf{2016}, \emph{144}, 114106\relax
\mciteBstWouldAddEndPuncttrue
\mciteSetBstMidEndSepPunct{\mcitedefaultmidpunct}
{\mcitedefaultendpunct}{\mcitedefaultseppunct}\relax
\EndOfBibitem
\bibitem[Richardson(2016)]{ric16a}
Richardson,~J.~O. Microcanonical and thermal instanton rate theory for chemical
  reactions at all temperatures. \emph{Faraday Disc.} \textbf{2016},
  \emph{195}, 49--67\relax
\mciteBstWouldAddEndPuncttrue
\mciteSetBstMidEndSepPunct{\mcitedefaultmidpunct}
{\mcitedefaultendpunct}{\mcitedefaultseppunct}\relax
\EndOfBibitem
\bibitem[Richardson(2018)]{ric18a}
Richardson,~J.~O. Ring-polymer instanton theory. \emph{Int. Rev. Phys. Chem.}
  \textbf{2018}, \emph{37}, 171--216\relax
\mciteBstWouldAddEndPuncttrue
\mciteSetBstMidEndSepPunct{\mcitedefaultmidpunct}
{\mcitedefaultendpunct}{\mcitedefaultseppunct}\relax
\EndOfBibitem
\bibitem[Seideman and Miller(1992)Seideman, and Miller]{mil91}
Seideman,~T.; Miller,~W.~H. Calculation of the cumulative reaction probability
  via a discrete variable representation with absorbing boundary conditions.
  \emph{J. Chem. Phys.} \textbf{1992}, \emph{96}, 4412--4422\relax
\mciteBstWouldAddEndPuncttrue
\mciteSetBstMidEndSepPunct{\mcitedefaultmidpunct}
{\mcitedefaultendpunct}{\mcitedefaultseppunct}\relax
\EndOfBibitem
\bibitem[Miller(1998)]{mil98a}
Miller,~W.~H. Direct and correct Calculation of Canonical and Microcanonical
  Rate Constants for Chemical Reactions. \emph{J. Phys. Chem. A} \textbf{1998},
  \emph{102}, 793--806\relax
\mciteBstWouldAddEndPuncttrue
\mciteSetBstMidEndSepPunct{\mcitedefaultmidpunct}
{\mcitedefaultendpunct}{\mcitedefaultseppunct}\relax
\EndOfBibitem
\bibitem[Miller(1998)]{mil98add}
Miller,~W.~H. Spiers Memorial Lecture Quantum and semiclassical theory of
  chemical reaction rates. \emph{Faraday Discuss.} \textbf{1998}, \emph{110},
  1--21\relax
\mciteBstWouldAddEndPuncttrue
\mciteSetBstMidEndSepPunct{\mcitedefaultmidpunct}
{\mcitedefaultendpunct}{\mcitedefaultseppunct}\relax
\EndOfBibitem
\bibitem[Takayanagi(1952)]{tak52}
Takayanagi,~K. On the Inelastic Collision between Molecules, II: Rotational
  Transition of H$_2$-molecule in the Collision with another H$_2$-molecule.
  \emph{Prog. Theor. Phys.} \textbf{1952}, \emph{8}, 497--508\relax
\mciteBstWouldAddEndPuncttrue
\mciteSetBstMidEndSepPunct{\mcitedefaultmidpunct}
{\mcitedefaultendpunct}{\mcitedefaultseppunct}\relax
\EndOfBibitem
\bibitem[Bowman(1991)]{bow91}
Bowman,~J.~M. Reduced dimensionality theory of quantum reactive scattering.
  \emph{J. Phys. Chem.} \textbf{1991}, \emph{95}, 4960--4968\relax
\mciteBstWouldAddEndPuncttrue
\mciteSetBstMidEndSepPunct{\mcitedefaultmidpunct}
{\mcitedefaultendpunct}{\mcitedefaultseppunct}\relax
\EndOfBibitem
\bibitem[Kryvohuz(2013)]{kry13}
Kryvohuz,~M. On the derivation of semiclassical expressions for quantum
  reaction rate constants in multidimensional systems. \emph{J. Chem. Phys.}
  \textbf{2013}, \emph{138}, 244114\relax
\mciteBstWouldAddEndPuncttrue
\mciteSetBstMidEndSepPunct{\mcitedefaultmidpunct}
{\mcitedefaultendpunct}{\mcitedefaultseppunct}\relax
\EndOfBibitem
\bibitem[McConnell \latin{et~al.}(2017)McConnell, L\"ohle, and
  K\"astner]{mcconnel17}
McConnell,~S.~R.; L\"ohle,~A.; K\"astner,~J. Rate constants from instanton
  theory via a microcanonical approach. \emph{J. Chem. Phys.} \textbf{2017},
  \emph{146}, 074105\relax
\mciteBstWouldAddEndPuncttrue
\mciteSetBstMidEndSepPunct{\mcitedefaultmidpunct}
{\mcitedefaultendpunct}{\mcitedefaultseppunct}\relax
\EndOfBibitem
\bibitem[Gutzwiller(1971)]{gut71}
Gutzwiller,~M.~C. Periodic Orbits and Classical Quantization Conditions.
  \emph{J. Math. Phys.} \textbf{1971}, \emph{12}, 343--358\relax
\mciteBstWouldAddEndPuncttrue
\mciteSetBstMidEndSepPunct{\mcitedefaultmidpunct}
{\mcitedefaultendpunct}{\mcitedefaultseppunct}\relax
\EndOfBibitem
\bibitem[Gutzwiller(1967)]{gut67}
Gutzwiller,~M.~C. Phase Integral Approximation in Momentum Space and the Bound
  States of an Atom. \emph{J. Math. Phys.} \textbf{1967}, \emph{8},
  1979--2000\relax
\mciteBstWouldAddEndPuncttrue
\mciteSetBstMidEndSepPunct{\mcitedefaultmidpunct}
{\mcitedefaultendpunct}{\mcitedefaultseppunct}\relax
\EndOfBibitem
\bibitem[Richardson \latin{et~al.}(2015)Richardson, Bauer, and Thoss]{ric15a}
Richardson,~J.~O.; Bauer,~R.; Thoss,~M. Semiclassical Green's functions and an
  instanton formulation of electron-transfer rates in the nonadiabatic limit.
  \emph{J. Chem. Phys.} \textbf{2015}, \emph{143}, 134115\relax
\mciteBstWouldAddEndPuncttrue
\mciteSetBstMidEndSepPunct{\mcitedefaultmidpunct}
{\mcitedefaultendpunct}{\mcitedefaultseppunct}\relax
\EndOfBibitem
\bibitem[Richardson(2015)]{ric15b}
Richardson,~J.~O. Ring-polymer instanton theory of electron transfer in the
  nonadiabatic limit. \emph{J. Chem. Phys.} \textbf{2015}, \emph{143},
  134116\relax
\mciteBstWouldAddEndPuncttrue
\mciteSetBstMidEndSepPunct{\mcitedefaultmidpunct}
{\mcitedefaultendpunct}{\mcitedefaultseppunct}\relax
\EndOfBibitem
\bibitem[M{\"u}ller and Brown(1979)M{\"u}ller, and Brown]{mul79}
M{\"u}ller,~K.; Brown,~L.~D. Location of saddle points and minimum energy paths
  by a constrained simplex optimization procedure. \emph{Theor. Chim. Acta}
  \textbf{1979}, \emph{53}, 75--93\relax
\mciteBstWouldAddEndPuncttrue
\mciteSetBstMidEndSepPunct{\mcitedefaultmidpunct}
{\mcitedefaultendpunct}{\mcitedefaultseppunct}\relax
\EndOfBibitem
\bibitem[McConnell and K\"astner(2017)McConnell, and K\"astner]{mcconnel171}
McConnell,~S.~R.; K\"astner,~J. Instanton rate constant calculations close to
  and above the crossover temperature. \emph{J. Comput. Chem.} \textbf{2017},
  \emph{38}, 2570--2580\relax
\mciteBstWouldAddEndPuncttrue
\mciteSetBstMidEndSepPunct{\mcitedefaultmidpunct}
{\mcitedefaultendpunct}{\mcitedefaultseppunct}\relax
\EndOfBibitem
\bibitem[Schreiner \latin{et~al.}(2011)Schreiner, Reisenauer, Ley, Gerbig, Wu,
  and Allen]{sch11a}
Schreiner,~P.~R.; Reisenauer,~H.~P.; Ley,~D.; Gerbig,~D.; Wu,~C.-H.;
  Allen,~W.~D. Methylhydroxycarbene: Tunneling Control of a Chemical Reaction.
  \emph{Science} \textbf{2011}, \emph{332}, 1300--1303\relax
\mciteBstWouldAddEndPuncttrue
\mciteSetBstMidEndSepPunct{\mcitedefaultmidpunct}
{\mcitedefaultendpunct}{\mcitedefaultseppunct}\relax
\EndOfBibitem
\bibitem[K\"astner(2013)]{kae13}
K\"astner,~J. The Path Length Determines the Tunneling Decay of Substituted
  Carbenes. \emph{Chem. Eur. J.} \textbf{2013}, \emph{19}, 8207--8212\relax
\mciteBstWouldAddEndPuncttrue
\mciteSetBstMidEndSepPunct{\mcitedefaultmidpunct}
{\mcitedefaultendpunct}{\mcitedefaultseppunct}\relax
\EndOfBibitem
\bibitem[Dirac(1929)]{dir29}
Dirac,~P. Quantum Mechanics of Many-Electron Systems. \emph{Proc. Royal Soc.
  (London) A} \textbf{1929}, \emph{123}, 714--733\relax
\mciteBstWouldAddEndPuncttrue
\mciteSetBstMidEndSepPunct{\mcitedefaultmidpunct}
{\mcitedefaultendpunct}{\mcitedefaultseppunct}\relax
\EndOfBibitem
\bibitem[Slater(1951)]{sla51}
Slater,~J. A simplification of the Hartree-Fock method. \emph{Phys. Rev.}
  \textbf{1951}, \emph{81}, 385--390\relax
\mciteBstWouldAddEndPuncttrue
\mciteSetBstMidEndSepPunct{\mcitedefaultmidpunct}
{\mcitedefaultendpunct}{\mcitedefaultseppunct}\relax
\EndOfBibitem
\bibitem[Vosko \latin{et~al.}(1980)Vosko, Wilk, and Nusair]{vos80}
Vosko,~S.~H.; Wilk,~L.; Nusair,~M. Accurate spin-dependent electron liquid
  correlation energies for local spin density calculations: a critical
  analysis. \emph{Can. J. Phys.} \textbf{1980}, \emph{58}, 1200--1211\relax
\mciteBstWouldAddEndPuncttrue
\mciteSetBstMidEndSepPunct{\mcitedefaultmidpunct}
{\mcitedefaultendpunct}{\mcitedefaultseppunct}\relax
\EndOfBibitem
\bibitem[Becke(1988)]{bec88}
Becke,~A.~D. Density-functional exchange-energy approximation with correct
  asymptotic behavior. \emph{Phys. Rev. A} \textbf{1988}, \emph{38},
  3098--3100\relax
\mciteBstWouldAddEndPuncttrue
\mciteSetBstMidEndSepPunct{\mcitedefaultmidpunct}
{\mcitedefaultendpunct}{\mcitedefaultseppunct}\relax
\EndOfBibitem
\bibitem[Lee \latin{et~al.}(1988)Lee, Yang, and Parr]{lee88}
Lee,~C.; Yang,~W.; Parr,~R.~G. Development of the Colle-Salvetti
  correlation-energy formula into a functional of the electron density.
  \emph{Phys. Rev. B} \textbf{1988}, \emph{37}, 785--789\relax
\mciteBstWouldAddEndPuncttrue
\mciteSetBstMidEndSepPunct{\mcitedefaultmidpunct}
{\mcitedefaultendpunct}{\mcitedefaultseppunct}\relax
\EndOfBibitem
\bibitem[Becke(1993)]{bec93}
Becke,~A.~D. Density-functional thermochemistry. III. The role of exact
  exchange. \emph{J. Chem. Phys.} \textbf{1993}, \emph{98}, 5648\relax
\mciteBstWouldAddEndPuncttrue
\mciteSetBstMidEndSepPunct{\mcitedefaultmidpunct}
{\mcitedefaultendpunct}{\mcitedefaultseppunct}\relax
\EndOfBibitem
\bibitem[Weigend and Ahlrichs(2005)Weigend, and Ahlrichs]{weigend2005}
Weigend,~F.; Ahlrichs,~R. Balanced basis sets of split valence{,} triple zeta
  valence and quadruple zeta valence quality for H to Rn: Design and assessment
  of accuracy. \emph{Phys. Chem. Chem. Phys.} \textbf{2005}, \emph{7},
  3297--3305\relax
\mciteBstWouldAddEndPuncttrue
\mciteSetBstMidEndSepPunct{\mcitedefaultmidpunct}
{\mcitedefaultendpunct}{\mcitedefaultseppunct}\relax
\EndOfBibitem
\bibitem[tur()]{turbomole}
{TURBOMOLE}, a development of {University of Karlsruhe} and {Forschungszentrum
  Karlsruhe GmbH}, 1989-2007, {TURBOMOLE GmbH}, since 2007; available from {\tt
  http://www.turbomole.com}.\relax
\mciteBstWouldAddEndPunctfalse
\mciteSetBstMidEndSepPunct{\mcitedefaultmidpunct}
{}{\mcitedefaultseppunct}\relax
\EndOfBibitem
\bibitem[K\"astner \latin{et~al.}(2009)K\"astner, Carr, Keal, Thiel, Wander,
  and Sherwood]{kae09a}
K\"astner,~J.; Carr,~J.~M.; Keal,~T.~W.; Thiel,~W.; Wander,~A.; Sherwood,~P.
  {DL-FIND}: an Open-Source Geometry Optimizer for Atomistic Simulations.
  \emph{J. Phys. Chem. A} \textbf{2009}, \emph{113}, 11856--11865\relax
\mciteBstWouldAddEndPuncttrue
\mciteSetBstMidEndSepPunct{\mcitedefaultmidpunct}
{\mcitedefaultendpunct}{\mcitedefaultseppunct}\relax
\EndOfBibitem
\bibitem[Metz \latin{et~al.}(2014)Metz, K\"astner, Sokol, Keal, and
  Sherwood]{met14}
Metz,~S.; K\"astner,~J.; Sokol,~A.~A.; Keal,~T.~W.; Sherwood,~P. ChemShell---a
  modular software package for QM/MM simulations. \emph{WIREs Comput. Mol.
  Sci.} \textbf{2014}, \emph{4}, 101--110\relax
\mciteBstWouldAddEndPuncttrue
\mciteSetBstMidEndSepPunct{\mcitedefaultmidpunct}
{\mcitedefaultendpunct}{\mcitedefaultseppunct}\relax
\EndOfBibitem
\bibitem[Eichkorn \latin{et~al.}(1997)Eichkorn, Weigend, Treutler, and
  Ahlrichs]{eichkorn1997}
Eichkorn,~K.; Weigend,~F.; Treutler,~O.; Ahlrichs,~R. Auxiliary basis sets for
  main row atoms and transition metals and their use to approximate Coulomb
  potentials. \emph{Theor. Chem. Acc.} \textbf{1997}, \emph{97}, 119--124\relax
\mciteBstWouldAddEndPuncttrue
\mciteSetBstMidEndSepPunct{\mcitedefaultmidpunct}
{\mcitedefaultendpunct}{\mcitedefaultseppunct}\relax
\EndOfBibitem
\end{mcitethebibliography}
	
	% URL and PMID removed via makefile: rah03
	% \bibliography{mod}
	
\newpage

\end{document}